\documentclass[lettersize,journal]{IEEEtran}
\usepackage{amsmath,amsfonts}
\usepackage{algorithmic}
\usepackage{array}
\usepackage[caption=false,font=normalsize,labelfont=sf,textfont=sf]{subfig}
\usepackage{textcomp}
\usepackage{stfloats}
\usepackage{url}
\usepackage{verbatim}
\usepackage{graphicx}
\usepackage{cite}
\def\BibTeX{{\rm B\kern-.05em{\sc i\kern-.025em b}\kern-.08em
    T\kern-.1667em\lower.7ex\hbox{E}\kern-.125emX}}
\usepackage{balance}
\usepackage{enumitem}
\usepackage{multirow}

\usepackage{siunitx, soul}
\usepackage{amsthm}
\usepackage{amsmath}
\usepackage{amssymb}
\usepackage{algorithm}
\usepackage{algorithmic}

\newtheorem{rem}{Remark}
\usepackage{booktabs}

\newcommand{\q}[1]{{\bf{#1}}}

\newcommand{\tr}{\text{Tr}}

\begin{document}
\bstctlcite{IEEEexample:BSTcontrol}

\title{Wideband Cognitive Radio for Joint Communication and Sensing: Optimization of  Subcarrier Allocation and beamforming}

\author{Diluka Galappaththige, \IEEEmembership{Member, IEEE}, and  Chintha Tellambura, \IEEEmembership{Fellow, IEEE}
\thanks{D. Galappaththige, and C. Tellambura with the Department of Electrical and Computer Engineering, University of Alberta, Edmonton, AB, T6G 1H9, Canada (e-mail: \{diluka.lg, ct4\}@ualberta.ca).} \vspace{-5mm}}

\markboth{Journal of \LaTeX\ Class Files,~Vol.~18, No.~9, September~2020}%
{How to Use the IEEEtran \LaTeX \ Templates}

\maketitle

\begin{abstract}
As data traffic grows, wireless systems shift to higher frequency bands (6 GHz and above), where radar systems also operate. This coexistence demands effective interference management and efficient wideband utilization. Cognitive Radio (CR) offers a solution but remains limited to single-node or narrowband systems. This paper introduces a generalized wideband CR-enabled communication and sensing system with multiple users and targets. We propose a communication and sensing sub-carrier allocations framework, followed by transmit beamforming for the primary communication BS and sensing signal design for the secondary radar BS. The goal is to maximize the communication sum rate while ensuring sensing requirements, minimizing interference, and adhering to power constraints. To solve the resulting non-convex problem, we develop a manifold optimization algorithm for communication-only sub-carriers and an alternating optimization approach using the generalized Rayleigh quotient and semidefinite relaxation for communication-sensing sub-carriers. Compared to a non-cooperative benchmark, the proposed system achieves a \qty{10}{\percent} gain in communication sum rate and a \qty{32}{\percent} gain in sensing sum rate with \num{12} BS antennas.
\end{abstract}


\section{Introduction}
\IEEEPARstart{T}{he} increasing demand for high data rates has shifted wireless communication systems towards higher frequency bands, such as millimeter-wave (mmWave), i.e., \qtyrange{30}{300}{\GHz}, where radar systems also operate \cite{Azar2024}. This spectral overlap necessitates the coexistence of radar and communication systems, requiring effective interference management and efficient wideband spectrum utilization.  The cognitive radio (CR) approach can address these challenges,  enabling integrated sensing and communication (ISAC) within the same spectral resources \cite{Khawar2014, Wang2019, Qian2021, Mao2024, xu2024,  Liu2024, Elfiatoure2024}. Leveraging dynamic spectrum access, CR facilitates the efficient operation of communication and sensing systems, optimizing spectral efficiency \cite{Zheng2019}.

CR is a software-defined radio technology that detects the RF environment and adapts parameters like frequency, power, and beamforming for optimal performance. It enables dynamic spectrum access and interference management, making it ideal for ISAC. Unlike static spectrum allocation, CR allows licensed primary and opportunistic secondary systems to coexist efficiently \cite{Hilal2023}. The secondary system performs spectrum sensing to identify available channels and minimize interference using energy detection, matched filtering, and cyclostationary feature detection \cite{Hilal2023}.

\subsection{Wideband CR for ISAC}
Although a few studies \cite{Khawar2014, Wang2019, Qian2021, Mao2024, xu2024, Liu2024, Elfiatoure2024} examine the CR approach for radar-communication coexistence (Section~\ref{sec_previous_work}), they have significant limitations. Most focus on single-user communication and single-target sensing, optimizing communication or sensing rather than both. However, effective coexistence requires enhancing both domains simultaneously.

Additionally, these studies largely overlook higher-frequency bands, particularly mmWave. By restricting their scope to narrowband systems—where communication and sensing operate over limited bandwidths with a single carrier—they fail to utilize the full spectral potential \cite{Rappaport2015}. Narrowband assumptions oversimplify real-world propagation by treating channels as flat-fading, ignoring the complexities of frequency-selective fading in wideband systems. In contrast, wideband systems span multiple subcarriers, supporting high-capacity communication and high-resolution sensing, making them far superior for ISAC \cite{Rappaport2015}.

Despite its advantages, CR for ISAC remains largely unexplored. The few existing studies \cite{Khawar2014, Wang2019, Qian2021, Mao2024, xu2024, Liu2024, Elfiatoure2024} are not sufficient to capture its potential. To our knowledge, this is the first study to introduce a wideband CR approach for ISAC, filling this gap and paving the way for future advancements.

Nevertheless, the wideband CR approach for radar-communication coexistence poses several technical challenges. These include:
\begin{enumerate}
    \item The primary system must efficiently allocate sub-carriers to optimize communication performance while preventing over-utilization of spectrum resources.

    \item The secondary sensing system must perform spectrum sensing to identify available spectral opportunities and select optimal sensing sub-carriers, minimizing or avoiding mutual interference between the primary and secondary systems.

    \item Adaptive beamforming is needed at both the primary and secondary systems based on sub-carrier utilization (i.e., communication-only or communication and sensing) and transmission protocols, ensuring optimal communication and sensing performance while minimizing mutual interference between the systems.

    \item Both the primary and secondary systems must support multiple access for users and multi-target detection, respectively, to ensure effective operation in a shared spectrum resource.
\end{enumerate}

\subsection{Our Contribution} Inspired by these challenges and the potential of wideband radar-communication coexistence, this study investigates a generalized wideband CR-enabled communication and sensing system with multiple users and targets (Fig.~\ref{fig_SystemModel}). A framework is developed for sub-carrier allocation for communication and selection for sensing. Based on this, optimal transmit beamforming and receiver combining at the communication and radar base stations (BSs) are designed to maximize the communication sum rate while ensuring the required sensing performance.

The main contributions of this paper are as follows:
\begin{enumerate}
    \item It proposes a CR approach to enable a wideband ISAC system. It consists of a primary communication system with a multiple antenna BS and multiple users and a secondary sensing system with a full-duplex (FD) BS and multiple targets is investigated. To our knowledge, this is the first study to address this system model. 

    \item To effectively utilize the spectrum, sub-carriers must be effectively allocated. To this end, the primary BS allocates subcarriers for user communication based on channel gains. At the same time, the secondary BS employs energy detection-based spectrum sensing to select sensing subcarriers, aiming to minimize mutual interference.

    \item Based on the sub-carrier assignment, the primary BS transmit beamforming ($\{\q{w}_{l,k}\}$), the secondary BS sensing signal ($\{\q{s}_l\}$), and secondary BS sensing combining ($\{\q{u}_{l,t}\}$) are optimized. The objective is to maximize the primary communication sum rate while meeting the sensing rate requirements for each target at the secondary BS, minimizing interference from the secondary system on primary users, and adhering to the transmit power constraints at the BSs.

    \item The proposed problem $(\mathcal{P})$ \eqref{prob_P} is non-convex due to involved product of optimization variables. For communication-only sub-carriers, a manifold optimization (MO)-based algorithm is developed. For communication and sensing sub-carries, an alternating optimization (AO) algorithm is proposed, leveraging the generalized Rayleigh quotient method and the semidefinite relaxation (SDR) technique.

    \item Convergence and complexity analyses and numerical examples are presented to evaluate the performance of the wideband radar-communication coexistence system. The proposed system is compared against non-cooperative radar and communication systems (i.e., no cooperation between the primary and secondary systems), as well as communication-only and sensing-only schemes. With a configuration of \num{12} antennas at both the primary and secondary BSs (for transmission and reception), the proposed cooperative design achieves a \qty{10.0}{\percent} gain in the sum communication rate and a \qty{32.2}{\percent} gain in the sum sensing rate.
\end{enumerate}

\subsection{Previous Contributions}\label{sec_previous_work}
A few works consider the CR approach for enabling communication and sensing \cite{Khawar2014, Wang2019, Qian2021, Mao2024, xu2024,  Liu2024, Elfiatoure2024}. Reference \cite{Khawar2014} studies a spatial approach utilizing a spectrum sharing between a multiple-input multiple-output (MIMO) radar and a cellular system with multiple BSs. The key idea is to project the radar signal onto the null space of the interference channels between the radar and cellular systems using an interference-channel-selection algorithm to minimize the radar's interference with communication. In \cite{Wang2019}, two communication-centric beamforming designs are proposed to facilitate coexistence between downlink multi-user communication and MIMO radar. These designs maximize the communication sum rate while constraining the communication interference at the radar. For a radar-communication CR system with a single user and a single-target radar, reference \cite{Qian2021} proposes an AO-based beamforming algorithm to minimize radar interference at the communication user.

In \cite{Mao2024}, a two-stage coexistence framework is investigated for detecting and tracking a radar system with a single target alongside multiple small-cell BSs. This work proposes AO-based algorithms to minimize the system's transmit power and enhance radar sensing performance during the detection and tracking stages while maintaining communication quality. Reference \cite{xu2024} explores the cognitive operation of a reconfigurable intelligent surface (RIS)-assisted primary communication system with a single-antenna BS and a secondary ISAC system. An AO algorithm, leveraging Dinkelbach’s transform and successive convex approximation (SCA), is proposed to maximize the targets' sensing signal-to-interference-plus-noise ratio (SINR). However, this approach does not consider the performance of the primary system. Reference \cite{Liu2024} also considers a dual active RIS-assisted primary communication system comprising a single-antenna BS, a single user, and a secondary MIMO radar system with a single target. It designs the radar's beamforming and the reflecting coefficients of the active RISs to maximize the communication rate, utilizing an AO algorithm based on penalty dual decomposition. In \cite{Elfiatoure2024}, the coexistence of a multi-user MIMO communication system and a single-target MIMO radar is analyzed. Closed-form expressions for communication rate and target detection probability are derived using conventional precoding, including maximum ratio (MR), zero-forcing (ZF), and protective ZF (where the information-bearing signal is projected onto the null space of the radar channel). A power control scheme is also proposed to maximize detection probability while ensuring per-user rate requirements.

\textit{Notation}: 
Vectors and matrices are denoted by boldface lowercase and uppercase letters.  $\mathbb{C}^{M\times N}$ and ${\mathbb{R}^{M \times 1}}$ represent $M\times N$  and $M\times 1$  complex and  real vectors.  For matrix $\mathbf{A}$, $\mathbf{A}^{\rm{H}}$ and $\mathbf{A}^{\rm{T}}$ are its Hermitian conjugate transpose and transpose. $[\q{A}]_{m,n}$ denotes the $\{m,n\}$-th element of matrix $\q{A}$. $\mathbf{I}_M$ and $\mathbf{0}_M$ are $M\times M$ identity and all-zero matrices. The Euclidean norm and absolute value operators are  $\|\cdot\|$ and $|\cdot|$. Expectation and trace operators are $\mathbb{E}\{\cdot\}$ and $\tr(\cdot)$.  The distribution of a circularly symmetric complex Gaussian (CSCG) random vector with mean $\boldsymbol{\mu}$ and covariance matrix $\mathbf{C}$ is denoted by $\sim \mathcal{CN}(\boldsymbol{\mu}, \mathbf{C})$. The operation $\text{unt}(\mathbf{a})=\left[\frac{a_1}{|a_1|}, \ldots, \frac{a_n}{|a_n|}\right]$. $\mathbf A \otimes  \mathbf B $ is the Hadamard product. Further, $\mathcal{O}$ expresses the big-O notation. 

\section{Preliminaries}\label{Sec_Prel}
This section outlines the system, channel, and signal models. Moreover, it also describes transmission protocol and the sub-carrier allocation for communication and sensing.

\begin{figure}[!t]\vspace{-0mm} 
    \centering 
    \def\svgwidth{230pt} 
    \fontsize{7}{7}\selectfont 
    \graphicspath{{Figures/}}
    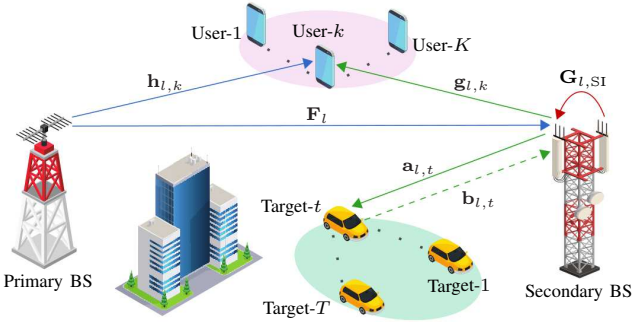 \vspace{-2mm} 
    \caption{A wideband CR-assisted communication and sensing system setup.}  \label{fig_SystemModel}\vspace{-2mm}
\end{figure}

\subsection{System Model}
A wideband CR approach is developed to enable communication and sensing. The system consists of a primary communication network with an HD BS serving \( K \) single-antenna users and a secondary radar-sensing network with an FD BS detecting \( T \) targets (Fig.~\ref{fig_SystemModel}). The primary BS has \( M \) uniform linear array (ULA) antennas, while the secondary BS has \( N \) transmitting and \( N \) receiving ULA antennas, all spaced at half-wavelengths \cite{Zhenyao2023}. The direct link between the primary BS and targets is assumed to be blocked or unavailable due to obstacles \cite{Liu2024}. This system could involve an outdoor BS communicating with mobile users while sensing smart devices indoors to track range, direction, and velocity for identification and environment mapping. However, mmWave signals struggle to penetrate buildings due to their short wavelengths and high susceptibility to absorption and reflection. This limits high-resolution indoor sensing, especially in NLoS scenarios, potentially requiring a separate sensing access point or BS \cite{Uwaechia2020}.

The primary BS transmits signals to users over the downlink sub-carriers. At the same time, the secondary BS performs energy detection-based spectrum sensing and selects sub-carriers with minimal interference from primary transmissions for target sensing. Let \( L \) denote the total set of sub-carriers. The primary BS allocates \( L_c (\leq L) \) sub-carriers per user for communication, whereas the secondary BS utilizes only \( L_s (\leq L) \) sub-carriers for sensing (Section~\ref{Sec_SC_alloc}).

\subsection{Channel Model}
A block flat-fading channel model is considered. In each fading block, at the $l$-th sub-carrier, $\q{h}_{l,k} \in \mathbb{C}^{M\times 1}$ for $k\in \{1, \dots, K\}$ is the channel between the primary BS and $k$-th user, $\q{F}_l \in \mathbb{C}^{M\times N}$ is the channel between the primary BS and the secondary BS receiver ULA, and $\q{g}_{l,k} \in \mathbb{C}^{N\times 1}$ is the channel between the secondary BS transmit ULA and $k$-th user. These pure communication channels are modeled as Rician fading and given as 
\begin{align}
    \q{a}  & = \sqrt{\frac{\kappa \zeta_{c}}{\kappa+1}} \q{a}^{\text{LoS}} + \sqrt{\frac{\zeta_{c}}{\kappa+1}} \q{a}^{\text{NLoS}}, \\
    \q{F}_l  & = \sqrt{\frac{\kappa \zeta_{F}}{\kappa+1}} \q{F}_l^{\text{LoS}} + \sqrt{\frac{\zeta_{F}}{\kappa+1}} \q{F}_l^{\text{NLoS}},
\end{align}
where $\q{c} \in \{\q{h}_{l,k}, \q{g}_{l,k}\}$, $\kappa$ is the Rician factor, and $\{\zeta_{c}, \zeta_{F}\}$ account for the large-scale path loss and shadowing, which stay constant for several coherence intervals. Moreover, $\q{a}^{\text{LoS}}$ and $\q{F}_l^{\text{LoS}}$ are the deterministic line-of-sight (LoS) components between the transmitter and receiver (i.e., modeled using array steering vectors in \eqref{eqn_array_steering}), and $\tilde{\q{c}} \sim \{\mathcal{CN}(\q{0}, \q{I}_{M}), \mathcal{CN}(\q{0}, \q{I}_{N})\}$ and $\tilde{\q{F}}_l \sim \mathcal{CN}(\q{0}_{M\times N}, \q{I}_{M} \otimes \q{I}_{N})$ are the NLoS components that follow the Rayleigh fading model.

On the other hand, following the echo signal representation in MIMO radar systems, the transmit/receiver channels between the secondary BS and targets, i.e., $\q{a}_{l,t}$ and $\q{b}_{l,t}$, are modeled as LoS channels \cite{Zhenyao2023}. The transmit/receiver array steering vectors to the direction $\theta_t$ at the $l$-th sub-carrier are thus modeled as
\begin{eqnarray}\label{eqn_array_steering}
    \bar{\q{c}} = \sqrt{\frac{1}{N}} \left[1, e^{j\pi \sin(\theta_t)}, \ldots, e^{j\pi (N-1) \sin(\theta_t)} \right]^{\rm{T}},
\end{eqnarray}
where $\bar{\q{c}} \in \{\q{a}_{l,t}, \q{b}_{l,t}\}$, $\theta_t$ is the $t$-th target's direction with respect to the $x$-axis of the coordinate system. Finally, $\q{G}_{l,\rm{SI}} \in  \mathbb{C}^{N\times N}$ is the SI channel matrix between the transmitter and the receiver antennas of the secondary BS and is modeled as a Rician fading channel with a Rician factor of  $\kappa_{\rm{SI}}$ \cite{Mohammadi2023, Diluka2024CFFD}. 

\subsection{Transmission Protocol}

\begin{figure}[!t]\vspace{-0mm} 
    \centering 
    \def\svgwidth{250pt} 
    \fontsize{7}{7}\selectfont 
    \graphicspath{{Figures/}}
    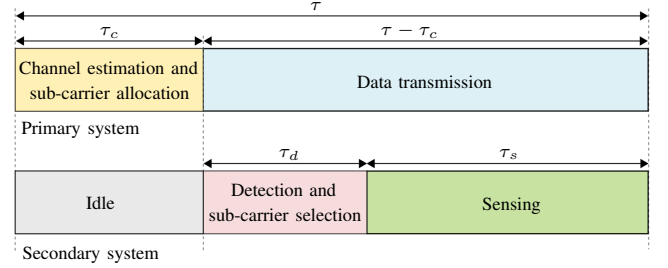 \vspace{-2mm} 
    \caption{Coherence time of the primary and secondary systems.}  \label{fig_CoheranceTime}\vspace{-2mm}
\end{figure}

The proposed system will use time-division duplex (TDD) transmission for both networks across all sub-carriers \cite{tse_viswanath_2005}. The primary network utilizes TDD for channel estimation, sub-carrier allocation, and data transmission (Fig.~\ref{fig_CoheranceTime}). Within each coherence block of length $\tau$, a portion of $\tau_c$ samples ($\tau_c < \tau$) is allocated for channel estimation and sub-carrier allocation. The remaining duration of the coherence block, $\tau - \tau_c$, is then dedicated to data transmission. On the other hand, the secondary network remains idle during the initial $\tau_c$ symbol periods (or it can estimate the primary user's channels, i.e., $\q{g}_{l,k}$). After that,  $\tau_d$ samples are allocated for spectrum sensing/detection and sub-carrier selection. The remaining $\tau_s = \tau-\tau_c-\tau_d$ symbol period is utilized for target sensing. 

\begin{figure}[!t]\vspace{-0mm} 
    \centering 
    \def\svgwidth{235pt} 
    \fontsize{8}{8}\selectfont 
    \graphicspath{{Figures/}}
    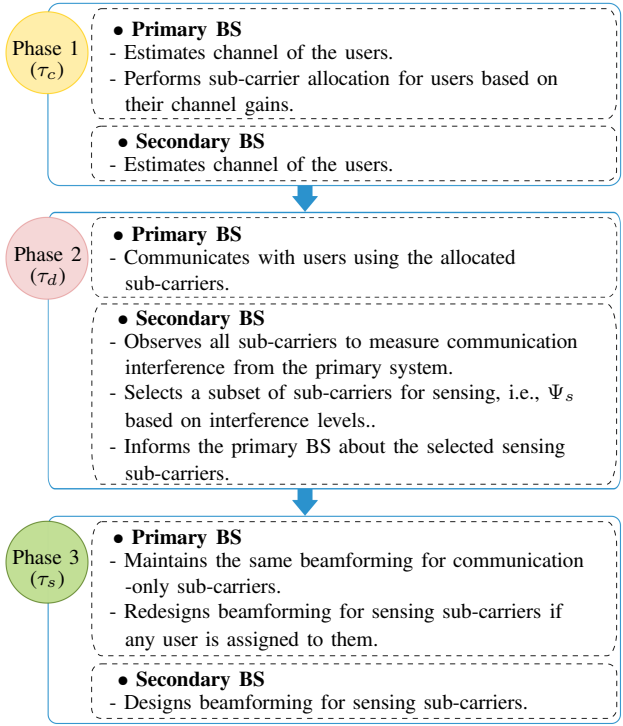 \vspace{-2mm} 
    \caption{The key parts of the proposed CR network.}  \label{fig_Flowchart}\vspace{-2mm}
\end{figure}

Fig.~\ref{fig_Flowchart}  illustrates the key processes in the proposed system.

\begin{rem}
The following standard assumptions are employed:
\begin{enumerate}[label=(\roman*)]
    \item In phase 1 (i.e., during $\tau_c$), the primary BS estimates user channels using uplink pilots, which the secondary BS can also leverage for primary user channel estimation. Well-established methods such as least squares (LS) and minimum mean squared error (MMSE) estimators facilitate accurate channel state information (CSI) acquisition \cite{Marzettabook2016, Nayebi2018}, ensuring the BSs and users have complete CSI knowledge.
    
    \item The secondary BS is assumed to have pre-estimated target angular directions, i.e., $\theta_{t}$, for beamforming, obtained from prior scanning \cite{Tsinos2021Joint, Zhenyao2023, Wu2018}. This prior knowledge enables more efficient beamforming design.
    
    \item The BSs are connected via a dedicated control link, which operates separately from communication links \cite{positioningLTE}. This link facilitates the exchange of essential system information, including CSI, beamforming weights, and synchronization commands \cite{positioningLTE}, ensuring effective coordination. The exchanged messages are typically lightweight and low-rate, minimizing overhead while maintaining seamless system operation.
\end{enumerate}
\end{rem}

\subsection{Sub-Carrier Allocation and Selection}\label{Sec_SC_alloc}
The primary BS leverages the estimated CSI to allocate sub-carriers for each user in phase 1 \cite{Liu2010}. This is achieved based on the channel power gains, ensuring efficient resource utilization and maximized throughput. Specifically, the BS evaluates the channel power gains across all available sub-carriers and assigns a subset of sub-carriers with the highest power gains to each user \cite{Liu2010}. We assume that each user is assigned with $L_c\geq 1 $ number of sub-carriers. To this end, the binary index  variable is defined as 
\begin{align}\label{eqn_SC_coefficient}
    \alpha_{l,k} = \begin{cases} 
    1, & \text{$l$-th sub-carrier  assigned to user $k$},\\
   0, & \text{otherwise}.
\end{cases}
\end{align}
Note that each sub-carrier can be assigned to multiple users, enabling sub-carrier sharing among users to enhance spectral efficiency and flexibility in resource allocation.

Conversely, during phase 2,  the secondary BS performs energy detection-based spectrum sensing. In particular, based on the primary received signal power at each sub-carrier, which acts as interference for sensing, the secondary BS selects a subset of sub-carriers for sensing. Specifically, it selects the least interference sub-carriers for sensing operation (Section~\ref{Sec_signal_mod}). This approach minimizes the impact of primary system operation on the secondary sensing performance, and improves target detection accuracy \cite{Liu2010}.

\subsection{Signal Model}\label{Sec_signal_mod}
Given the sub-carrier allocation, the primary BS transmits communication signal $\q{x}_l \in \mathbb{C}^{M\times 1}$ for $l\in\{1,\ldots,L\}$ to the users. This signal at the $l$-th sub-carrier is thus given as
\begin{align}
    \q{x}_l =  \sum\nolimits_{k =1}^{K} \alpha_{l,k} \q{w}_{l,k} q_{k},
\end{align}
where $q_{k} \in\mathbb{C}$ is the intended data symbol for the $k$-th user with unit power, i.e., $\mathbb{E}\{\vert q_{k} \vert^2 \} = 1$, $\q{w}_{l,k} \in \mathbb{C}^{M\times 1}$ is the primary BS data beamforming vector for the $k$-th user at the $l$-th sub-carrier, and $\alpha_{l,k}$ is the sub-carrier allocation coefficient in \eqref{eqn_SC_coefficient}. The received signal at the $k$-th user at the $l$-th sub-carrier is given by
\begin{align}\label{eqn_rx_user_k}
    y_k &= \q{h}_{l,k}^{\rm{H}} \q{x}_l + z_{l,k} \nonumber \\
    &= \alpha_{l,k} \q{h}_{l,k}^{\rm{H}} \q{w}_{l,k} q_k + \sum\nolimits_{i\neq k}^{K} \alpha_{l,i} \q{h}_{l,k}^{\rm{H}} \q{w}_{l,i} q_i + z_{l,k},
\end{align}
where $z_{l,k} \sim \mathcal{CN}(0,\sigma^2)$ is the $k$-th user additive white Gaussian noise (AWGN) at the $l$-th sub-carrier. 

During $\tau_d$, the secondary BS performs energy detection-based spectrum sensing to select the least communication interference sub-carriers for target sensing. The received primary communication signal at the secondary BS at the $l$-th sub-carrier is given as
\begin{align}
    \q{y}_{l,b} = \q{F}_l^{\rm{H}} \q{x}_l + \q{z}_{l,b},
\end{align}
where $\q{z}_{l,b} \sim \mathcal{CN}(\q{0}, \sigma^2\q{I}_N)$ is the AWGN at the secondary BS. The interference energy in the $l$-th sub-carrier at the secondary BS is thus given as $I_{l,b} = \Vert \q{y}_{l,b} \Vert^2$. To minimize it across all sub-carriers, the secondary BS selects a subset $\Psi_s$ $(\vert \Psi_s \vert = L_s)$ with the lowest $I_{l,b}$ terms,  where $L_s$ is the number of sub-carriers selected for sensing. This subset is defined as 
\begin{align}
    \Psi_s = \text{argmin}_{\{1,\ldots, L\}} I_{l,b}. 
\end{align}
As this minimizes the primary communication interference on secondary sensing performance,  sensing accuracy will improve. Thus, $\Psi_s$ identifies the sub-carriers that provide cleaner channels for target detection. 

\begin{rem}
    Although interference-based sub-carrier selection provides a simple and intuitive approach to improve sensing accuracy, it may not be the optimal strategy \cite{Gouissem2016}. More advanced selection methods may enhance performance by considering additional factors such as spectral correlation, signal sparsity, or statistical learning-based interference estimation \cite{Gouissem2016}. Future research can explore optimization-driven approaches like convex optimization and mixed-integer programming, machine learning techniques such as reinforcement learning and deep neural networks for adaptive sub-carrier selection, or game-theoretic models for dynamic resource allocation.
\end{rem}

Once the set of sensing sub-carriers is selected, the secondary BS transmits sensing signal $\q{s}_l \in \mathbb{C}^{N\times 1}$ with the covariance matrix $\q{S}_l \triangleq \mathbb{E}\{\q{s}_l \q{s}_l^{\rm{H}}\}$ for $l\in \Psi_s$ to perform target sensing \cite{Zhenyao2023}. Then, the secondary BS processes the target echo, i.e., reflected signal from the target, to extract the target's state information \cite{Zhenyao2023}. The received signal at the secondary BS in the $l$-th sub-carrier, i.e., $\q{y}_{l,s} \in \mathbb{C}^{N\times 1}$, is given as
\begin{align}\label{eqn_rx_BS}
    \!\! \q{y}_{l,s} = \sum\nolimits_{t=1}^{T} \beta_t \q{b}_{l,t} \q{a}_{l,t}^{\rm{H}} \q{s}_l + \q{F}_l^{\rm{H}} \q{x}_l + \sqrt{\beta_{\rm{SI}}} \q{G}_{l,\rm{SI}}^{\rm{H}} \q{s}_l  + \q{z}_{l,s},
\end{align}
where $\q{z}_{l,s} \sim \mathcal{CN}(\q{0},\sigma^2 \q{I}_N)$ is the AWGN vector at the secondary BS in the $l$-th sub-carrier and $\beta_t \q{b}_{l,t} \q{a}_{l,t}^{\rm{H}} \q{s}_l$ is the $t$-th target reflection, where $\beta_t \in \mathbb{C}$ is the complex amplitude of target reflection, accounting for the round-trip path loss and the radar cross-section (RCS) of the target \cite{Liu2022}.  Specifically, path loss accounts for signal attenuation over distance, whereas RCS determines how much power is reflected toward the BS based on the target's size, shape, and materials. It is also assumed that BS uses clutter rejection techniques to minimize the reflected clutter interference from the surrounding environment \cite{James2010RadarBook}. 

In \eqref{eqn_rx_BS}, the second term is the interference from the primary transmission and the third term is the SI at the secondary BS receiver, resulting from  FD transmission and reception, and $0 < \beta_{\rm{SI}} \ll 1$ is a constant that quantifies the SI cancellation ability of the secondary FD BS~\cite{Mohammadi2023}. Without loss of generality, we assume imperfect SI cancellation at the BS. The secondary BS then applies the sensing combiner, $\q{u}_{l,t} \in \mathbb{C}^{N\times 1}$ for $l \in \Psi_s$ and $t\in \{1,\ldots, T\}$,  to  the received echo signal \eqref{eqn_rx_BS} to capture the desired reflected signal of the $t$-th target. The post-processed signal for obtaining $t$-th target's sensing information at the $l$-th sub-carrier is given as
\begin{align}\label{eqn_rx_Sens}
    y_{l, t} &= \beta_t \q{u}_{l,t}^{\rm{H}} \q{b}_{l,t} \q{a}_{l,t}^{\rm{H}} \q{s}_l + \sum\nolimits_{j\neq t}^{T} \beta_j \q{u}_{l,t}^{\rm{H}} \q{b}_{l,j} \q{a}_{l,j}^{\rm{H}} \q{s}_l \nonumber\\
    &+ \q{u}_{l,t}^{\rm{H}} \q{F}_l^{\rm{H}} \q{x}_l + \sqrt{\beta_{\rm{SI}}} \q{u}_{l,t}^{\rm{H}} \q{G}_{l,\rm{SI}}^{\rm{H}} \q{s}_l + \q{u}_{l,t}^{\rm{H}} \q{z}_{l,s}. 
\end{align}

Suppose that the secondary BS uses the $l$-th sub-carrier for sensing. Thus, during the sensing phase $\tau_s$, the secondary transmissions interfere with primary communication on the same $l$-th sub-carrier. Assuming the $k$-th user is active on the $l$-th sub-carrier, the  received signal during $\tau_s$ is given by
\begin{align}\label{eqn_rx_user_k_sens}
    y_k' = \alpha_{l,k} \q{h}_{l,k}^{\rm{H}} \q{w}_{l,k} q_k + \sum\nolimits_{i\neq k}^{K} \alpha_{l,i} \q{h}_{l,k}^{\rm{H}} \q{w}_{l,i} q_i + \q{g}_{l,k}^{\rm{H}} \q{s}_l +z_{l,k}. 
\end{align}
Note that the sub-carriers selected for sensing, i.e., $l\in \Psi_s$, the primary BS must design two distinct beamforming vectors: one during the detection/selection phase $\tau_d$ and $\tau_s$ and another during the sensing phase $\tau_s$. These beamforming vectors are crucial for effectively serving the users assigned to the selected sub-carriers based on the respective received signals at the users, i.e., \eqref{eqn_rx_user_k} and \eqref{eqn_rx_user_k_sens}.

\section{Communication and Sensing Performance}\label{Sec_performance}
The CR system performance is determined by the communication rates of the users and the targets' sensing rate at the secondary BS.

\subsection{Communication Performance}
The users utilize the received signal from the primary BS to decode their intended information. The rate of the $k$-th user at the $l$-th sub-carrier can be approximated by
\begin{align}
    \mathcal{R}_{l,k}^{\rm{Com}} \approx \begin{cases}
        \frac{\tau_d}{\tau}\log_2(1+ \gamma_{l,k}) + \frac{\tau_s}{\tau} &\!\!\!\! \log_2(1+ \gamma_{l,k}'), \nonumber \\
        & \text{if} \quad l\in \Psi_s,\\
        \frac{\tau-\tau_c}{\tau}\log_2(1+ \gamma_{l,k}), & \text{otherwise},
    \end{cases}
\end{align}
where $\gamma_{l,k}$ and $\gamma_{l,k}'$ are the received SINR at the $k$-th user and defined by using \eqref{eqn_rx_user_k} and \eqref{eqn_rx_user_k_sens}, respectively, as
\begin{align}\label{eqn_gamma}
    \gamma_{l,k} &= \frac{\alpha_{l,k}^2 \vert \q{h}_{l,k}^{\rm{H}} \q{w}_{l,k}\vert^2}{\sum_{i\neq k}^{K} \alpha_{l,i}^2 \vert \q{h}_{l,k}^{\rm{H}} \q{w}_{l,i} \vert^2 + \sigma^2}, \\
    \gamma_{l,k}' &= \frac{\alpha_{l,k}^2 \vert \q{h}_{l,k}^{\rm{H}} \q{w}_{l,k}\vert^2}{\sum_{i\neq k}^{K} \alpha_{l,i}^2 \vert \q{h}_{l,k}^{\rm{H}} \q{w}_{l,i} \vert^2 + \q{g}_{l,k}^{\rm{H}} \q{S}_l \q{g}_{l,k} + \sigma^2}.
\end{align}

\subsection{Sensing Performance}
The transmit beampattern gain and the mean squared error (MSE) of the transmit beampattern are widely used sensing performance measures \cite{Stoica2007}. However, these metrics do not account for the receiver's beam pattern or multi-target interference, which can introduce ambiguities in multi-target detection due to signal interference from multiple reflections \cite{Zhenyao2023, Cui2014}.

In contrast, the  Cram\'{e}r-Rao bound (CRB) focuses solely on the lower bound of estimation error (i.e., accuracy) \cite{Tang2019, Cui2014}. While CRB quantifies the precision of parameter estimation (e.g., angle, distance, velocity), it does not capture how much environmental information is accumulated over time.

To address these limitations, sensing SINR or sensing rate has been proposed as a performance metric \cite{Cui2014, Zhenyao2023}. Notably, the target detection probability is proportional to its sensing SINR or rate, facilitating target detection by incorporating both transmit and receive beamforming \cite{Cui2014, Zhenyao2023}.

Due to its benefits,  the sensing rate is employed to measure sensing performance. From \eqref{eqn_rx_Sens}, the sensing rate of the $t$-th target in the $l$-th sub-carrier at the secondary BS is given as
\begin{align}
    \mathcal{R}_{l,t}^{\rm{Sen}} \approx \begin{cases}
        \frac{\tau_s}{\tau}{\rm{log}}_2(1 + \Upsilon_{l,t}), & \text{if} \quad l\in \Psi_s,\\
        0, & \text{otherwise},
    \end{cases}
\end{align}
where $\Upsilon_{l,k}$ is the sensing SINR of the $t$-th target at the $l$-th sub-carrier and given in \eqref{eqn_sens_SINR_target},
\begin{figure*}
 \begin{align}\label{eqn_sens_SINR_target}
        \Upsilon_{l,k} &= \frac{\vert\beta_t\vert^2 \mathbb{E} \left\{ \vert \q{u}_{l,t}^{\rm{H}} \q{b}_{l,t} \q{a}_{l,t}^{\rm{H}} \q{s}_l \vert^2 \right\}}{ \sum\nolimits_{j\neq t}^{T} \vert\beta_j\vert^2 \mathbb{E} \left\{ \vert \q{u}_{l,t}^{\rm{H}} \q{b}_{l,t} \q{a}_{l,t}^{\rm{H}} \q{s}_l \vert^2 \right\}  + \mathbb{E} \left\{ \vert \q{u}_{l,t}^{\rm{H}} \q{F}_{l}^{\rm{H}} \q{x}_l \vert^2 \right\}  + \beta_{\rm{SI}}  \mathbb{E} \left\{ \vert \q{u}_{l,t}^{\rm{H}} \q{G}_{l,\rm{SI}}^{\rm{H}} \q{s}_l \vert^2 \right\} + \mathbb{E} \left\{ \vert \q{u}_{l,t}^{\rm{H}} \q{z}_{l,s} \vert^2 \right\}   }\nonumber\\
        &= \frac{\vert \beta_t\vert^2 \q{u}_{l,t}^{\rm{H}} \q{b}_{l,t} \q{a}_{l,t}^{\rm{H}} \q{S}_l \q{b}_{l,t}^{\rm{H}} \q{a}_{l,t} \q{u}_{l,t} }{ \q{u}_{l,t}^{\rm{H}}\left( \sum\nolimits_{j\neq t}^{T} \vert\beta_j\vert^2  \q{b}_{l,j} \q{a}_{l,j}^{\rm{H}} \q{S}_l \q{b}_{l,j}^{\rm{H}} \q{a}_{l,j} +  \q{F}_{l}^{\rm{H}} \q{R}_{x,l} \q{F}_{l} + \beta_{\rm{SI}}   \q{G}_{l,\rm{SI}}^{\rm{H}} \q{S}_l \q{G}_{l,\rm{SI}} + \sigma^2 \q{I}_N \right) \q{u}_{l,t} }
 \end{align}
\hrulefill

\vspace{-2mm}

\end{figure*}
where $\q{R}_{x,l} \triangleq \mathbb{E} \{\q{x}_l \q{x}_l^{\rm{H}} \} =  \sum_{k=1}^{K} \q{w}_{l,k} \q{w}_{l,k}^{\rm{H}}$ is the primary BS transmitted signal covariance matrix at the $l$-th sub-carrier \cite{Zhenyao2023}.

\section{Problem Formulation}
The primary objective is to maximize the sum rate, i.e., communication and sensing sum rate for $l\in \Psi_s$ and communication sum rate for $l\in\{1,\ldots, L\}\setminus \Psi_s$. In particular, for each sub-carrier, this goal is achieved by jointly optimizing the primary BS transmit beamforming $\{\q{w}_{l,k}\}$, the secondary BS sensing covariance matrix $\{\q{S}_{l}\}$, and the secondary BS sensing combining $\{\q{u}_{l,t}\}$. The optimization problem for the $l$-th sub-carrier is thus formulated as follows: 
\begin{subequations}\label{prob_P}
\begin{align}
    (\mathcal{P}):~& \max_{\mathcal{A}_l} \quad  \sum\nolimits_{k=1}^{K} \mathcal{R}_{l,k}^{\rm{Com}} +  \vartheta_l \sum\nolimits_{t=1}^{T} \mathcal{R}_{l,t}^{\rm{Sen}}, \label{prob_P_obj}  \\
    \text{s.t.} \quad &   \vert \q{g}_{l,k}^{\rm{H}} \q{s}_l \vert^2 \leq \delta_{\rm{max}},\quad \text{if}~l\in\Psi_s, \label{prob_P_interference}\\
    & \sum\nolimits_{i=1}^{K} \alpha_{l,i}^2 \Vert \q{w}_{l,i} \Vert^2 \leq p_{\rm{max}},\quad \forall l, \label{prob_P_prim_tx_pow} \\
    & \Vert \q{s}_{l} \Vert^2 \leq p_{\rm{max}}',\quad \text{if}~l \in\Psi_s,\label{prob_P_sec_tx_pow} \\
    & \Vert \q{u}_{l,t} \Vert^2 = 1,\quad \text{if}~l\in\Psi_s,\label{prob_P_sec_comb} 
\end{align}
\end{subequations}
where $\vartheta_l = 1$ if $l\in\Psi_s$ and $\vartheta_l = 0$ otherwise, and the set of optimization variables at the $l$-th sub-carrier is defined as
\begin{align}
    \mathcal{A}_l = \begin{cases}
        \{\q{w}_{l,k}, \q{S}_{l}, \q{u}_{l,t} \}, & \text{if} \quad l\in\Psi_s,\\
        \{\q{w}_{l,k}\}, & \text{otherwise}.
    \end{cases}
\end{align}

In $(\mathcal{P})$, the constraint \eqref{prob_P_interference} limits the secondary sensing interference on the primary user with maximal allowable interference power $\delta_{\rm{max}}$, constraints \eqref{prob_P_prim_tx_pow} and \eqref{prob_P_sec_tx_pow} set the primary and secondary BS transmit powers with maximum allowable transmit powers $p_{\rm{max}}$ and $p_{\rm{max}}'$, respectively, and constraint \eqref{prob_P_sec_comb} is the normalization constraint for the secondary BS sensing combiners. 

\section{Proposed Solution}
This solves  $(\mathcal{P})$ based on the sub-carrier utilization, i.e., communication-only or communication and sensing. 

\subsection{Communication-only Beamforming}
For the sub-carries $l\in\{1,\ldots, L\}\setminus \Psi_s$, problem $(\mathcal{P})$ becomes the primary BS beamforming problem. It is thus reformulated as the following equivalent problem:
\begin{subequations}\label{prob_P1}
\begin{align}
    (\mathcal{P}1):~& \max_{\q{w}_{l,k}} \quad \frac{\tau-\tau_c}{\tau} \sum\nolimits_{k=1}^{K} \log_2(1+ \gamma_{l,k}), \label{prob_P1_obj}  \\
    \text{s.t.} \quad &  \sum\nolimits_{i=1}^{K}\alpha_{l,i}^2 \Vert \q{w}_{l,i} \Vert^2 \leq p_{\rm{max}}. \label{prob_P1_prim_tx_pow} 
\end{align}
\end{subequations}
Note that the problem $(\mathcal{P}1)$ is non-convex due to the non-convex objective function. Hence, to address this, we utilize fractional programming (FP) and MO to obtain the optimal primary BS transmit beamforming vectors \cite{zargari2024riemannian, zargari2024CFISAC}. 

However, $(\mathcal{P}1)$ cannot be directly tackled by the MO as the optimization variable involves separate $\q{w}_{l,k}$. Thus, we first introduce a matrix $\q{V}_l = [\alpha_{l,1}\q{w}_{l,1}, \ldots, \alpha_{l,K}\q{w}_{l,K}]$ and equivalent transformations are performed on $(\mathcal{P}1)$ to solve it with MO. Moreover, we also define an index matrix $\q{E} = \q{I}_K \in \mathbb{R}^{K\times K}$ to select the corresponding beamforming vectors of a particular user, i.e., the primary BS beamforming corresponding to the $k$-th user can be thus represented as $\q{w}_{l,k} = \q{V}_l \q{E}_k$, where $\q{E}_{k}$ is the $k$-th column of $\q{E}$. Thereby, $(\mathcal{P}1)$ can be equivalently represented as 
\begin{subequations}\label{prob_P2}
\begin{align}
    (\mathcal{P}2):~& \max_{\q{V}_{l}} \quad \frac{\tau-\tau_c}{\tau} \sum\nolimits_{k=1}^{K} \log_2(1+ \bar{\gamma}_{l,k}), \label{prob_P2_obj}  \\
    \text{s.t.} \quad &  \tr(\q{V}_l \q{V}_l^{\rm{H}}) \leq p_{\rm{max}}. \label{prob_P2_prim_tx_pow} 
\end{align}
\end{subequations}
where 
\begin{align}\label{eqn_gamma_bar}
    \bar{\gamma}_{l,k}= \frac{ \vert \q{h}_{l,k}^{\rm{H}} \q{V}_{l} \q{E}_k \vert^2}{\sum_{i\neq k}^{K}  \vert \q{h}_{l,k}^{\rm{H}} \q{V}_{l} \q{E}_i \vert^2 + \sigma^2}.
\end{align}
In $(\mathcal{P}2)$, to address the challenging sum-log terms in the objective, we invoke the FP technique. In particular, the Lagrangian dual transform is utilized to move $\bar{\gamma}_{l,k}$ to the outside of $\log_2\left(1 +  \bar{\gamma}_{l,k} \right)$. This converts the original problem into an equivalent version, where $\q{V}_l$ is a solution to $(\mathcal{P}2)$ only if it is also a solution to equivalent problem $(\mathcal{P}3)$  \cite[\textit{Theorem 3}]{Shen2018FPpart2}. Consequently, an auxiliary variable vector $\boldsymbol{\mu}_l = [\mu_{l,1}, \ldots, \mu_{l,K}]$ is introduced to replace the each SINR term in \eqref{prob_P2_obj} such that $\mu_{l,k} \leq \bar{\gamma}_{l,k}$. Then, $(\mathcal{P}2)$ is reformulated as \cite{Shen2018} 
\begin{subequations}\label{prob_P3}
\begin{align}
    (\mathcal{P}3):~& \max_{\q{v}_l, \boldsymbol{\mu}_l} ~~f(\q{V}_l, \boldsymbol{\mu}_l) = \frac{\tau-\tau_c}{\tau \ln(2)} \sum\nolimits_{k = 1}^{K} \ln(1 + \mu_{l,k})  \nonumber\\
    &+ \frac{\tau-\tau_c}{\tau \ln(2)} \sum\nolimits_{k=1}^{K} \left( - \mu_{l,k} + \frac{(1 + \mu_{l,k})\bar{\gamma}_{l,k}}{1 + \bar{\gamma}_{l,k}} \right), \label{prob_P3_obj}\\
    \text{s.t} \quad & \eqref{prob_P2_prim_tx_pow}.
\end{align}
\end{subequations}
Problem $(\mathcal{P}3)$ can be considered as a two-part optimization problem: (i) an outer optimization over $\q{V}_l$ with fixed $\boldsymbol{\mu}_l$ and (ii) an inner optimization over $\boldsymbol{\mu}_l$ with fixed $\q{V}_l$ \cite{Shen2018}. To address $(\mathcal{P}3)$, the variables $\q{V}_l$ and $\boldsymbol{\mu}_l$ are alternately optimized until the objective function converges \cite{zargari2024riemannian, zargari2024CFISAC}.  

\subsubsection{Optimizing $\boldsymbol{\mu}_l$} 
For a given $\q{V}_l$, the objective $f(\q{V}_l, \boldsymbol{\mu}_l)$ becomes a concave and differentiable function with respect to $\boldsymbol{\mu}_l$. Thus, the optimal $\boldsymbol{\mu}_l$ can be obtained by setting each $\frac{\partial f(\q{V}_l, \boldsymbol{\mu}_l)}{\partial \mu_{l,k}}$ to zero. Accordingly, the optimal $\mu_{l,k}$ is given by $\mu_{l,k}^* = \bar{\gamma}_{l,k}$ for $k\in\{1,\dots, K\}$. Note that substituting $\boldsymbol{\mu}_l^*$ back into $f(\q{V}_l, \boldsymbol{\mu}_l)$ recovers the exact sum-of-logarithms objective function in $(\mathcal{P}2)$. 

\subsubsection{Optimizing $\q{V}_l$}
For a given $\boldsymbol{\mu}_l$, the objective function in \eqref{prob_P3_obj} can be simplified by eliminating the constant terms with respect to $\q{V}_l$. As a result, $(\mathcal{P}3)$ can be reformulated as follows:
\begin{subequations}\label{prob_p4}
\begin{align}
    \!\!\!\!(\mathcal{P}4):~& \max_{\q{V}_l} ~ f(\q{V}_l) = \sum_{k=1}^{K} \frac{\hat{\mu}_{l,k}  \vert  \q{h}_{l,k}^{\rm{H}} \q{V}_l \q{E}_{k}\vert^2}{\sum_{i=1}^{K}  \vert \q{h}_{l,k}^{\rm{H}} \q{V}_l \q{E}_{i} \vert^2 + \sigma^2}, \label{prob_P4_obj}\\
    \text{s.t} \quad & \eqref{prob_P2_prim_tx_pow},
\end{align}
\end{subequations}
where $\hat{\mu}_{l,k} = 1+\mu_{l,k}$ for $k\in\{1, \ldots, K\}$. Note that the problem $(\mathcal{P}4)$ and the original problem $(\mathcal{P}1)$ are equivalent, and transformations do not degrade performance.

\begin{rem}\label{rem_equivalance}
The equivalence between $(\mathcal{P}1)$ and $(\mathcal{P}4)$ can be established as follows: As $\q{w}_{l,k} = \q{V}_l \q{E}_k$,  $(\mathcal{P}2)$ is identical to $(\mathcal{P}1)$. In $(\mathcal{P}3)$, substituting optimal $\boldsymbol{\mu}_l^*$ back in $f(\q{V}_l, \boldsymbol{\mu}_l)$ recovers the original sum-of-logarithms in the objective function in $(\mathcal{P}2)$, i.e., $\frac{\tau-\tau_c}{\tau}\sum_{k=1}^{K} \log_2\left(1 +  \bar{\gamma}_{l,k} \right)$, exactly. This establishes the equivalency between $(\mathcal{P}2)$ and $(\mathcal{P}3)$  \cite{Shen2018, Shen2018FPpart2}. For a given $\boldsymbol{\mu}_l$, the only term that depends on $\q{V}_l$ in \eqref{prob_P3} is $\sum_{k=1}^{K} \frac{(1 + \mu_{l,k})\bar{\gamma}_{l,k}}{1 + \bar{\gamma}_{l,k}}$, and the constant terms with respect to $\q{V}_l$ can be eliminated \cite{boyd2004convex}. Hence, the objective and the constraint in $(\mathcal{P}3)$ and $(\mathcal{P}4)$ are the same, establishing their equivalence. Therefore, the above equivalences prove the equivalence between the initial problem $(\mathcal{P}1)$ and the final version $(\mathcal{P}4)$ \cite{Shen2018, Shen2018FPpart2, boyd2004convex}.
\end{rem}

Problem $(\mathcal{P}4)$ can be efficiently solved via the MO technique. First, a modified matrix $\tilde{\q{V}}_l = [\q{v}_{l,1}, \ldots, \q{v}_{l,K}]$ is introduced by normalizing the power constraint \eqref{prob_P2_prim_tx_pow}, such that $\tr(\tilde{\q{V}}_l \tilde{\q{V}}_l^{\rm{H}}) = \tr(\q{V}_l \q{V}_l^{\rm{H}}) + ||\q{n}_l||_2^2 = 1$, where $\q{v}_{l,k} = [\alpha_{l,k}\q{w}_{l,k}^{\rm{T}}, n_{l,k}]^{\rm{T}}$ for $k \in \{1,\ldots, K\}$, and $\q{n}_l = [n_{l,1}, \ldots, n_{l,K}]$ is an auxiliary vector introduced to simplify power normalization while preserving the constraint. This normalization results in a complex sphere manifold $\mathcal{M} = \{ \tilde{\q{V}}_l \in \mathbb{C}^{(M+1) \times (K)} \:|\: \tr(\tilde{\q{V}}_l \tilde{\q{V}}_l^{\rm{H}}) = 1 \}$. Therefore, $(\mathcal{P}4)$ is transformed into an unconstrained optimization problem on $\mathcal{M}$ as follows:
\begin{align}\label{prob_P5}
  \!\!\!(\mathcal{P}5):~ \min_{\tilde{\q{V}}_l \in \mathcal{M}}~ f(\tilde{\q{V}}_l) = - \sum_{k=1}^{K} \frac{\hat{\mu}_{l,k}  \vert  \hat{\q{h}}_{l,k}^{\rm{H}} \tilde{\q{V}}_l \q{E}_{k}\vert^2}{\sum_{i=1}^{K}  \vert \hat{\q{h}}_{l,k}^{\rm{H}} \tilde{\q{V}}_l \q{E}_{i} \vert^2 + \sigma^2},
\end{align}
where $\hat{\q{h}}_{l,k} = \sqrt{p_{\rm{max}}}[\q{h}_{l,k}, 0]$ is adjusted to match the problem's dimensionality and scaling. The optimization variable $\q{V}_l$ is constrained to lie on $\mathcal{M}$, aligning with the MO framework. \textbf{Algorithm~\ref{alg_MO}} provides the framework for optimizing $(\mathcal{P}5)$ on $\mathcal{M}$, involving the following key steps \cite{liu2020simple, zargari2024riemannian, zargari2024CFISAC}:

\begin{algorithm}[!t] 
\caption{: Communication-only Beamforming Algorithm}
\label{alg_MO}
\begin{algorithmic}[1]
\STATE \textbf{Initialization}: Initial point $\tilde{\q{V}}_{l,0} \in \mathcal{M}$, convergence tolerance $\delta_1>0$ and $\delta_2>0$, and set $r_1=0$.
\WHILE{$\text{dist}(f(\tilde{\q{V}}_{l,r_1}), f(\tilde{\q{V}}_{{l,r_1 + 1}})) \geq \delta_2$}
    \STATE Update $\boldsymbol{\eta}_{l,0} = -{\rm{grad}}_{\tilde{\q{V}}_{l,0}} f(\tilde{\q{V}}_l)$ and set $r=0$.
    \WHILE{$\|{\rm{grad}}_{\tilde{\q{V}}_{l,r}} f(\tilde{\q{V}}_l)\|_2 > \delta_1$}
        \STATE Calculate Armijo backtracking line search step  $\varrho_{l,r}$.
        \STATE Update $\tilde{\q{V}}_{l,r+1}$ using the retraction $R_{\tilde{\q{V}}_{l,r}}(\varrho_{l,r}\boldsymbol{\eta}_{l,r})$.
        \STATE Update $\mathcal{T}_{\tilde{\q{V}}_{l, r} \rightarrow \tilde{\q{V}}_{l,r+1}}(\boldsymbol{\eta}_{l,r})$.
        \STATE Compute the Hestenes-Stiefel parameter $\nu_{l,r}$.
        \STATE Update the search direction $\boldsymbol{\eta}_{l,r+1}$.
        \STATE $r \leftarrow r + 1$.
    \ENDWHILE
    \STATE ${r_1} \leftarrow {r_1} + 1$.
    \STATE $\tilde{\q{V}}_{l,0} \leftarrow \tilde{\q{V}}_{l,r+1}$.
\ENDWHILE
\STATE \textbf{Output}: $\q{V}_l^* = \tilde{\q{V}}_l^{*}(1:M, :)$.
\end{algorithmic} 
\end{algorithm}

\textbf{Gradient computation:} This step computes the Riemannian gradient of $f(\tilde{\q{V}}_l)$ on $\mathcal{M}$. This is achieved by projecting the Euclidean gradient onto the tangent space $T_{\tilde{\q{V}}_{l,r}}\mathcal{M}$ at the current point $\tilde{\q{V}}_{l,r}$. The Euclidean gradient of $f(\q{V})$, i.e., $\nabla_{\tilde{\q{V}}_{l,r}} f(\tilde{\q{V}}_l)$, is given by \eqref{derivtive_eq}.
\begin{figure*}[!t]\vspace{-0mm}
\begin{align}  \label{derivtive_eq}
    \nabla_{\tilde{\q{V}}_{l,r}} f(\tilde{\q{V}}_l) & =  
    \sum_{k=1}^{K} -\hat{\mu}_{l,k}  \left(\frac{2  \hat{\q{h}}_{l,k}^{\rm{H}} \tilde{\q{V}}_{l,r} \q{E}_{k} \hat{\q{h}}_{l,k} \q{E}_{k}^{\rm{H}} }{\sum_{j=1}^{K}  \vert \hat{\q{h}}_{l,k}^{\rm{H}} \tilde{\q{V}}_{l,r} \q{E}_{j}\vert^2 + \sigma^2} -   \sum_{i=1}^{K} \frac{2  \vert\hat{\q{h}}_{l,k}^{\rm{H}} \tilde{\q{V}}_{l,r} \q{E}_{k}\vert^2  \hat{\q{h}}_{l,k}^{\rm{H}} \tilde{\q{V}}_{l,r} \q{E}_{i} \hat{\q{h}}_{l,k}\q{E}_{i}^{\rm{H}}  }{\left(\sum_{j=1}^{K}  \vert \hat{\q{h}}_{l,k}^{\rm{H}} \tilde{\q{V}}_{l,r} \q{E}_{j}\vert^2 + \sigma^2\right)^2} \right) 
\end{align}

\vspace{-5mm}

\end{figure*}

\textbf{Search direction:} This step determines the search direction by choosing a descent direction in $T_{\tilde{\q{V}}_{l,r}}\mathcal{M}$. It can be given by $\boldsymbol{\eta}_{l, r+1} = -{\rm{grad}}_{\tilde{\q{V}}_{l, r+1}} f(\tilde{\q{V}}_l) + \nu_{l,r} \mathcal{T}_{\tilde{\q{V}}_{l,r} \rightarrow \tilde{\q{V}}_{l, r+1}}(\boldsymbol{\eta}_{l,r})$, where $\boldsymbol{\eta}_{l,r}$ is the current search direction and $\nu_{l,r}$ is computed using the Hestenes-Stiefel approach \cite{Shewchuk1994}.

\textbf{Retraction (Mapping):} This step applies a retraction operation, $R_{\tilde{\q{V}}_{l,r}} (\varrho_{l,r}\eta_{l,r}) = \text{unt} (\varrho_{l,r}\eta_{l,r})$, where $\varrho_{l,r}$ is the step size to map the updated point, which lies in the tangent space, back onto $\mathcal{M}$. This ensures that the next iterate remains on the manifold after the update. Interested readers are referred to \cite{liu2020simple, zargari2024riemannian, zargari2024CFISAC} and related literature for more insights and algorithmic details.

\subsection{Communication and Sensing Beamforming}
For sub-carriers \( l \in \Psi_s \), problem \( (\mathcal{P}) \) becomes a joint primary and secondary beamforming design problem with \( \mathcal{A}_l = \{\q{w}_{l,k}, \q{S}_{l}, \q{u}_{l,t} \} \) and \( \vartheta_l = 1 \). It is non-convex due to the product of optimization variables in the objective function. To address this, an AO strategy decouples the problem into two sub-problems, solving them alternately while keeping the other fixed \cite{bezdek2003convergence}. The process repeats until a stopping condition is met, making optimization more manageable and efficient.

\subsubsection{Sub-problem 1: Optimizing $\{\q{u}_{l,t}\}$}\label{Sec_receiver_combine}
For fixed $\{\q{w}_{l,k}, \q{S}_{l}\}$, the sensing rate $\mathcal{R}_{l,t}^{\rm{Sen}}$ is the only term that depends on $\{\q{u}_{l,t}\}$ in the objective function \eqref{prob_P_obj}. On the other hand, the sensing rate $\mathcal{R}_{l,t}^{\rm{Sen}}$ is a monotonically increasing function of its argument, i.e., the sensing SINR $\Upsilon_{l,t}$. Hence, we first replace the sensing rate with the corresponding seining SINR. Thereby, using the unique structure of the sensing SINR for each target in \eqref{eqn_sens_SINR_target}, this sub-problem can be transformed into a generalized Rayleigh quotient problem, providing closed-form optimal combiner vectors \cite{Stanczak2008book}. To this end, problem $(\mathcal{P})$ is transformed into the following optimization problem:
\begin{subequations}\label{prob_Q1} 
\begin{align}
   (\mathcal{Q}1):~& \max_{\{\q{u}_{l,t}\}}  \quad   \frac{\q{u}_{l,t}^{\rm{H}} \q{f}_{l,t} \q{f}_{l,t}^{\rm{H}} \q{u}_{l,t}}{\q{u}_{l,t}^{\rm{H}} \q{Q}_{l,t} \q{u}_{l,t}}, \\
   \text{s.t} \quad &  \Vert \q{u}_{l,t} \Vert^2 = 1,
\end{align}
\end{subequations}
where  $\q{f}_{l,t} = \vert\beta_t \vert \q{b}_{l,t} \q{a}_{l,t}^{\rm{H}} \q{s}_l$, and $ \q{Q}_{l,t} = \sum_{j\neq t}^{T} \vert\beta_j\vert^2  \q{b}_{l,j} \q{a}_{l,j}^{\rm{H}} \q{S}_l \q{b}_{l,j}^{\rm{H}} \q{a}_{l,j} +  \q{F}_{l}^{\rm{H}} \q{R}_{x,l} \q{F}_{l} + \beta_{\rm{SI}}   \q{G}_{l,\rm{SI}}^{\rm{H}} \q{S}_l \q{G}_{l,\rm{SI}} + \sigma^2 \q{I}_N$. Problem $(\mathcal{Q}1)$ is a generalized Rayleigh ratio quotient problem \cite{Stanczak2008book}. The optimal sensing combiner is thus given by
\begin{align}\label{eqn_opt_u}
    \q{u}_{l,t}^* = \frac{\q{Q}_{l,t}^{-1} \q{f}_{l,t}}{\Vert \q{Q}_{l,t}^{-1} \q{f}_{l,t} \Vert}, \quad \forall t,
\end{align}
which is a minimal mean-squared error (MMSE) filter \cite{Stanczak2008book}.

\subsubsection{Sub-problem 2: Optimizing $\{\q{w}_{l,k}, \q{S}_l\}$}\label{Sec_tx_beamforming}
For $\{\q{u}_{l,t}\}$, problem $(\mathcal{P})$ becomes the following joint transmit beamforming design at the primary and the secondary BSs. 
\begin{subequations}\label{prob_Q2}
\begin{align}
    \!(\mathcal{Q}2)\!:& \max_{\{\q{w}_{l,k}, \q{S}_l\}} \frac{\tau_s}{\tau} \sum_{k=1}^{K} \log_2(1+ \gamma_{l,k}') + \frac{\tau_s}{\tau} \sum_{t=1}^{T} {\rm{log}}_2(1 + \Upsilon_{l,t}), \label{prob_Q2_obj}  \\
    \text{s.t.} \quad &   \eqref{prob_P_interference}- \eqref{prob_P_sec_tx_pow}. 
\end{align}
\end{subequations}
In $(\mathcal{Q}2)$, owing to the interference terms within the communication and sensing SINRs, the objective function \eqref{prob_Q2_obj} is non-convex. To solve this, we employ the SDR technique. We first define the matrix $\q{W}_{l,k} \triangleq \q{w}_{l,k} \q{w}_{l,k}^{\rm{H}}$, where $\q{W}_{l,k}$ is semi-definite matrix with rank one constraint, i.e., ${\rm{Rank}}(\q{W}_{l,k}) = 1$. Then, utilizing SDR techniques to relax the highly non-convex rank one constraint, the resultant problem can be formulated as follows:
\begin{subequations}\label{prob_Q3}
\begin{align}
    (\mathcal{Q}3):~& \max_{\{\q{W}_{l,k}, \q{S}_l\}} \quad \frac{\tau_s}{\tau} \sum\nolimits_{k=1}^{K} \digamma_{l,k} + \frac{\tau_s}{\tau} \sum\nolimits_{t=1}^{T} \Phi_{l,t}, \label{prob_Q3_obj}  \\
    \text{s.t.} \quad & \tr\left(\q{g}_{l,k} \q{g}_{l,k}^{\rm{H}} \q{S}_{l} \right)  \leq \delta_{\rm{max}}, ~\forall k, \label{prob_Q3_interference}\\
    & \sum\nolimits_{i=1}^{K} \alpha_{l,i}^2 \tr\left(\q{W}_{l,i} \right) \leq p_{\rm{max}}, \label{prob_Q3_tx_pow} \\
    & \tr\left(\q{S}_{l} \right) \leq p_{\rm{max}}', \label{prob_Q3_sec_tx_pow} \\
    & \q{W}_{l,k},  \q{S}_l \succeq 0,~\forall k,
\end{align}
\end{subequations}
As \eqref{prob_Q2_obj} is not a convex function of the optimization variables, we utilize the SCA method to linearize the objective function, and $\digamma_{l,k}$ and $\Phi_{l,t}$ are given in \eqref{eqn_digamma} and  \eqref{eqn_phi}, respectively, 
\begin{figure*}
\begin{eqnarray}
    \digamma_{l,k} \!\!&=&\!\! \log_2 \left( \sum\nolimits_{i=1}^{K} \alpha_{l,i}^2 \tr(\q{h}_{l,k} \q{h}_{l,k}^{\rm{H}} \q{W}_{l,i}) + \tr(\q{g}_{l,k} \q{g}_{l,k}^{\rm{H}} \q{S}_{l}) + \sigma^2 \right) - \log_2(D_{l,k}) - \frac{ \tr(\q{g}_{l,k} \q{g}_{l,k}^{\rm{H}} (\q{S}_{l} - \q{S}_{l}^{(p)}))}{\ln(2) D_{l,k}}  \nonumber \\
    && -\sum\nolimits_{i\neq k}^{K} \frac{ \alpha_{l,i}^2 \tr(\q{h}_{l,k} \q{h}_{l,k}^{\rm{H}} (\q{W}_{l,i} - \q{W}_{l,i}^{(p)}))}{\ln(2) D_{l,k}} \label{eqn_digamma} \\
    \Phi_{l,t} \!\!&=& \!\! \log_2 \left( \sum\nolimits_{j=1}^{T} \vert\beta_j \vert^2  \tr(\q{f}_{l,tj} \q{f}_{l,tj}^{\rm{H}} \q{S}_{l}) + \sum\nolimits_{i=1}^{K} \alpha_{l,i}^2 \tr(\q{g}_{l,t}^{\rm{I}} (\q{g}_{l,t}^{\rm{I}})^{\rm{H}} \q{W}_{l,i}) + \beta_{\rm{SI}} \tr(\q{g}_{l,t}^{\rm{SI}} (\q{g}_{l,t}^{\rm{SI}})^{\rm{H}} \q{S}_{l})  +\sigma^2 \Vert \q{u}_{l,t}\Vert^2\right) -\log_2(B_{l,t}) \nonumber \\
    && \!\!\!\!\!\!\!\!\!\!\!\!\!\! \!\!\!\!\!\! - \sum\nolimits_{j\neq t}^{T} \frac{ \vert\beta_j \vert^2 \tr(\q{f}_{l,tj} \q{f}_{l,tj}^{\rm{H}} (\q{S}_{l} - \q{S}_{l}^{(p)}))}{\ln(2) B_{l,t}} - \sum\nolimits_{i=1}^{K} \frac{ \alpha_{l,i}^2 \tr(\q{g}_{l,t}^{\rm{I}} (\q{g}_{l,t}^{\rm{I}})^{\rm{H}} (\q{W}_{l,i} - \q{W}_{l,i}^{(p)}))}{\ln(2) B_{l,t}} - \frac{ \beta_{\rm{SI}} \tr(\q{g}_{l,t}^{\rm{SI}} (\q{g}_{l,t}^{\rm{SI}})^{\rm{H}} (\q{S}_{l} - \q{S}_{l}^{(p)}))}{\ln(2) B_{l,t}} \label{eqn_phi} \quad
\end{eqnarray}
\hrulefill

\vspace{-2mm}

\end{figure*}
where $\q{f}_{l,tj} = \q{a}_{l,j} \q{b}_{l,j}^{\rm{H}} \q{u}_{l,t}$, $\q{g}_{l,t}^{\rm{I}} = \q{F}_l \q{u}_{l,t}$, and $\q{g}_{l,t}^{\rm{SI}} = \q{G}_{l, \rm{SI}} \q{u}_{l,t}$. Moreover, in \eqref{eqn_digamma} and  \eqref{eqn_phi}, $D_{l,k}$  and $B_{l,t}$ are defined as
\begin{align}
    D_{l,k} &\triangleq \sum\nolimits_{i\neq k}^{K} \alpha_{l,i}^2 \tr(\q{h}_{l,k} \q{h}_{l,k}^{\rm{H}} \q{W}_{l,i}^{(p)}) + \tr(\q{g}_{l,k} \q{g}_{l,k}^{\rm{H}} \q{S}_{l}^{(p)}) + \sigma^2, \\
    B_{l,t} &\triangleq  \sum\nolimits_{j\neq t}^{T} \vert\beta_j \vert^2  \tr(\q{f}_{l,tj} \q{f}_{l,tj}^{\rm{H}} \q{S}_{l}^{(p)}) + \beta_{\rm{SI}} \tr(\q{g}_{l,t}^{\rm{SI}} (\q{g}_{l,t}^{\rm{SI}})^{\rm{H}} \q{S}_{l}^{(p)}) \nonumber \\
    &+ \sum\nolimits_{i=1}^{K} \alpha_{l,i}^2 \tr(\q{g}_{l,t}^{\rm{I}} (\q{g}_{l,t}^{\rm{I}})^{\rm{H}} \q{W}_{l,i}^{(p)}) + \sigma^2 \Vert \q{u}_{l,t}\Vert^2,
\end{align}
where $(\cdot)^{(p)}$ denotes the previous iteration values of respective variables. This relaxed problem $(\mathcal{Q}3)$ is a standard semi-definite programming (SDP) problem and can be solved using the CVX Matlab tool \cite{boyd2004convex}. 

If the SDR solution satisfies ${\rm{Rank}}(\q{W}_{l,k}) =1$, the optimal transmit beamformers at the primary BS are obtained by eigenvalue decomposition \cite{Luo2010}. Let the eigenvalue decomposition of $\q{W}_{l,k}$ to be $\q{W}_{l,k} = \q{U}_{l,k} \boldsymbol{\Sigma}_{l,k} \q{U}_{l,k}^H$ where $\q{U}_{l,k}$ is a unitary matrix and $\boldsymbol{\Sigma}_{l,k} = \text{diag}(\lambda_{l,k1}, \dots, \lambda_{l,kM})$ is a diagonal matrix, both sized $M \times M$. If $\q{W}_{l,k}^*$ is rank one, the optimal transmit beamformer, $\q{w}_{l,k}^*$, is the eigenvector for the maximum eigenvalue. Otherwise, Gaussian randomization (GR) is employed to obtain a near-optimal solution for $(\mathcal{P}3)$ \cite{Luo2010}. However, satisfying the rank-one constraint by using GR  may cause a slight rate degradation $(\mathcal{P}2)$. Algorithm \ref{alg_beamforming} summarizes the steps to find the solution to $\mathcal{P}1$.

\begin{algorithm}[!t]
\caption{: Communication and Sensing Beamforming}
\begin{algorithmic}[1]
\label{alg_beamforming}
\STATE \textbf{Input}: Set the iteration counter $p = 0$, the convergence tolerance $\epsilon > 0$, initial feasible solution $\{ \q{w}_{l,k} \q{S}_l \}$. Initialize the objective function value $F^{(0)} = 0$.  
\WHILE{$ \frac{F^{(p+1)} - F^{(p)}}{F^{(p+1)}} \geq \epsilon$}
\STATE Solve $(\mathcal{Q}1)$ \eqref{prob_Q1} for $\q{u}_{l,t}^{(p+1)}$.
\STATE Begin - CVX.
\STATE Solve the convex problem $(\mathcal{Q}3)$ in \eqref{prob_Q3}.
\STATE End - CVX.
\STATE EVD  $\q{S}_l^*$ as $\q{S}_l^*=\q{P}_l \boldsymbol{\Lambda}_{l} \q{P}_l^H$, where $\q{P}=[\mathbf{p}_{l,1}, \dots, \q{p}_{l, N}]$.
\STATE \textbf{return}: $\q{s}_l^*= \q{p}_{l,1}$.
\FOR{$k=\{1, \dots, K\}$}
\STATE EVD  $\q{W}_{l, k}^*$ as $\q{W}_{l,k}^*=\q{U}_{l,k} \boldsymbol{\Sigma}_{l,k} \q{U}_{l,k}^H$, where $\q{U}_{l,k}=[\q{q}_{l,k1}, \dots, \q{q}_{l,kM}]$.
\IF{$\rm{Rank}(\q{W}_{l,k}^*)=1$,} 
    \STATE \textbf{return}: $\q{w}_{l,k}^*= \q{q}_{l,k1}$.
\ELSE
    \FOR{$d=1,\ldots,D$}
        \STATE Generate random  $\q{r}_{l,kd} = \q{U}_{l,k} \boldsymbol{\Sigma}_{l,k}^{1/2} \q{e}_{l,kd}$, where $\q{e}_{l,kd}  \sim \mathcal{CN}(\q{0}, \q{I}_{M})$.
        \STATE Check if $(\mathcal{Q}3)$ is feasible with $\q{r}_{l, kd}$. 
    \ENDFOR
    \STATE \textbf{return}: $\q{w}_{l,k}^*= \q{r}_{l,k}$, where  $\q{r}_{l,k}= \underset {d=\{1,\ldots,D\}}{\rm{arg \,min}} \,\, \q{r}_{l,kd}$.
\ENDIF 
\ENDFOR
\STATE Calculate the objective function value $F^{(p+1)}$.
\STATE Set $p\leftarrow p+1$;
\ENDWHILE
\STATE \textbf{Output}: Optimal solutions $\mathcal{A}_l^*$.
\end{algorithmic}
\end{algorithm}

\subsection{Computational Complexity and Algorithm Convergence}
\subsubsection{Algorithm \ref{alg_MO}}
The computational complexity of Algorithm \ref{alg_MO} is primarily due to the iterative process of the MO framework. In particular, the per-iteration complexity can be approximated as $\mathcal{O}(MK+M K^3)$. Let the number of iterations for convergence be $R$. Then, the total complexity can be approximated as $\mathcal{O}(R (MK+M K^3))$ \cite{liu2020simple, zargari2024riemannian, zargari2024CFISAC}.  

At each iteration, Algorithm \ref{alg_MO} generates a candidate solution $f(\tilde{\q{V}}_{l, r+1}) \leq f(\tilde{\q{V}}_{l, r}) + \epsilon_r$ with an infinite sequence $\{\epsilon_r\}$ that converges to zero, yielding a global minimizer for $(\mathcal{P}1)$. This monotonically decreasing nature and the upper constraint imposed on the objective function ensures convergence \cite{zargari2024riemannian}. 
\begin{proof}
   Please see Appendix~\ref{apdx_proof_Al_1}.
\end{proof}

\subsubsection{Algorithm \ref{alg_beamforming}}
The computational complexity of Algorithm \ref{alg_beamforming} depends on the two-sub problems. 
\begin{itemize}
    \item \textit{Sub-problem 1}: The Rayleigh quotient process involves computing the inverse of the matrix $\q{Q}_{l,t}$, requiring $\mathcal{O}(N^3)$. The MMSE filter for the $T$ targets adds complexity of $\mathcal{O}(TN^2)$. Thus, the total complexity for this sub-problem is $\mathcal{O}(TN^2 + N^3)$.

    \item \textit{Sub-problem 2}: The SDP sub-problem is solved via the interior-point method. From \cite[Theorem 3.12]{polik2010interior}, the complexity for a SDP problem with $m$ SDP constraints which includes a $n \times n$ positive semi-definite (PSD) matrix is given by $\mathcal{O}\left( \sqrt{n} \log\left(\frac{1}{\epsilon}\right) (mn^3 + m^2n^2 + m^3) \right)$, where $\epsilon > 0$ is the solution accuracy. For problem $(\mathcal{Q}3)$, with $n = M = N$ and $m = 3(K+T) + 2$, the computational complexity for solving $(\mathcal{Q}3)$ can be approximated as $\mathcal{O}\left( (K+T)M^3 \sqrt{M} \log\left(\frac{1}{\epsilon}\right) \right)$.
\end{itemize}
The overall computational complexity of Algorithm \ref{alg_beamforming} can be asymptotically given  as $\mathcal{O} \left(I_o \left((T + N) N^2 + (K+T)M^3 \sqrt{M} \log\left(\frac{1}{\epsilon}\right) \right) \right)$ where $I_o$ is the overall number of iterations to converge.

For the sub-carries $l\in\{1,\ldots, L\}$, problem $(\mathcal{P})$ is solved via the AO technique, yielding a local solution for each associated sub-problem. The AO method has well-established convergence \cite{bezdek2003convergence}. In particular, if the individual sub-problems converge, the overall problem also converges \cite{bezdek2003convergence}. Here, the SDR technique is used to optimize $\{\q{w}_{l,k}, \q{S}_l\}$ while $\{\q{u}_{l,t}\}$ is directly obtained via the Rayleigh ratio quotient method. The SDR is an established method with guaranteed convergence \cite{Luo2010, so2007approximating}, ensuring the overall convergence of Algorithm \ref{alg_beamforming}. Moreover, our simulation results validate this claim (Fig.~\ref{fig_Convergence_combined}).

\section{Simulation Results}
These evaluate the performance of the proposed wideband CR system to enable communication and sensing (Fig.~\ref{fig_SystemModel}). 

\subsection{Simulation Setup and Parameters}
The 3GPP urban micro (UMi) model is used to model the path loss values $\{\zeta_{c}, \zeta_F, \zeta_{\bar{c}}\}$ with the operating frequency of $f_c=\qty{28}{\GHz}$ \cite[Table 7.4.1-1]{3GPP2024}. The AWGN variance, $\sigma^2$, is modeled as $\sigma^2=10\log_{10}(N_0 B N_f)$ \qty{}{\dB m}, where $N_0=\qty{-174}{\dB m/\Hz}$, $B$ represents the bandwidth, $N_f$ denotes the noise figure. Unless otherwise specified, Table \ref{table_parameters} provides the simulation parameters \cite{zargari2024riemannian}. Each simulation point is averaged over \num{e3} iterations. 

\begin{table}[t]
\centering
\caption{Simulation and algorithm parameters.}\vspace{-2mm}
\label{table_parameters}
\begin{tabular}{c c c c}    
\hline
\textbf{Parameter}& \textbf{Value} & \textbf{Parameter}& \textbf{Value}\\  \hline \hline
$B$ & \qty{10}{\MHz}  & $\tau_d$  &  \num{100}\\ 
$N_f$ & \qty{10}{\dB} &  $\tau_s$  &  $\tau-\tau_d-\tau_c$ \\  
$M=N$ & \num{8}  & $\{p_{\rm{max}}, p_{\rm{max}}'\}$   & \qty{30}{\dB m}    \\ 
$L$ & \num{5}  &   $|\beta_t|$  & \num{e-2}  \\
$L_c$ & \num{3}  &  $\{\kappa, \kappa_{\rm{SI}}\}$ &  \qty{3}{\dB}   \\
$L_s$ & \num{1}  &  $\beta_{\rm{SI}}$ & \qty{-70}{\dB}   \\
$K$ & \{\num{3}, \num{5}\} &  $\delta_{\rm{max}}$ & \qty{-10}{\dB m}  \\ 
$T$ & \{\num{2}, \num{4}\}  &   $\{\delta_1,\delta_2\}$  & \num{e-6}   \\
$\tau$  &  \num{400} &  $\epsilon$  & \num{e-3}  \\  
$\tau_c$  &  $K$ &  $D$  & \num{e5}  \\  \hline
\end{tabular}
\vspace{-3mm}
\end{table}

The primary and secondary BSs are placed at $\{0,0\}$ and $\{70,0\}$. The users and targets are randomly distributed within circular regions centered at $\{30,0\}$ and $\{80,0\}$, respectively, with a radius of \qty{10}{\m} \cite{zargari2024riemannian}.

\subsection{Benchmark Schemes}
The following benchmarks are compared against the proposed system, which is labeled as ``WB-cognitive".

\subsubsection{Non-cooperative scheme}
The primary and secondary systems operate independently. The primary BS computes communication beamforming vectors using Algorithm~\ref{alg_MO}, while the secondary BS randomly selects $L_s$ sensing sub-carriers without considering primary users. 

\subsubsection{Communication-only scheme}This benchmark (`Comm-only') excludes the secondary BS and targets, establishing a baseline for communication-only performance. It aids in assessing trade-offs in the cognitive operation of both systems, i.e., communication and sensing.   

\subsubsection{Sensing-only scheme}
This benchmark (legend `Sens-only')  has the secondary sensing system without the primary communication system. 

Note that our Algorithm~\ref{alg_MO} and/or Algorithm~\ref{alg_beamforming} can accommodate all these benchmarks as special cases. 

\subsection{Convergence Rates of Algorithms}

\begin{figure}[!t]\vspace{-2mm}
    \centering
    \includegraphics[width=0.45\textwidth]{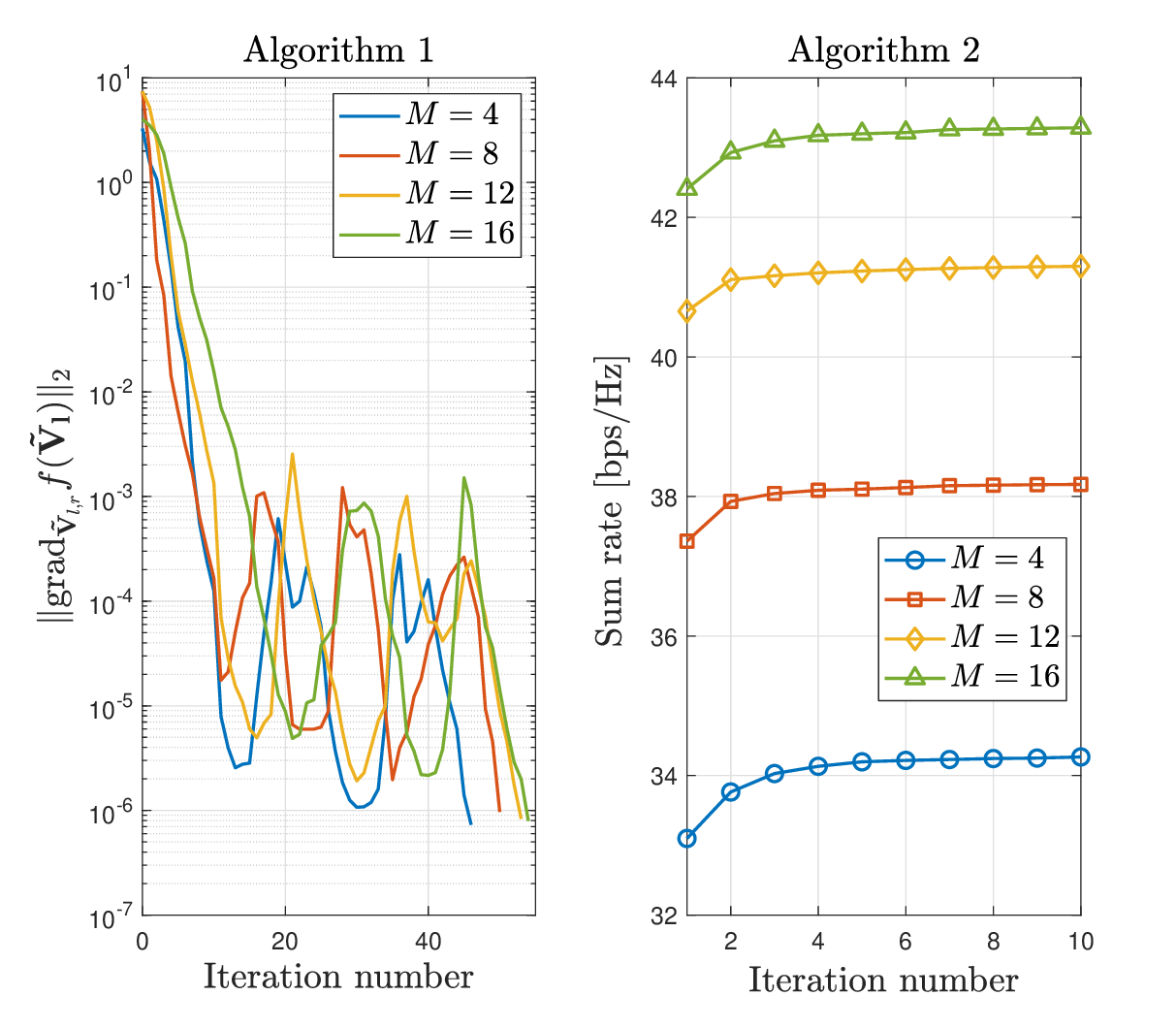}
    \vspace{-2mm}
    \caption{Convergence rates of Algorithm~\ref{alg_MO} (left) and Algorithm~\ref{alg_beamforming} (right) with different numbers of BS antennas.}
    \label{fig_Convergence_combined} \vspace{-2mm}
\end{figure}

Fig.~\ref{fig_Convergence_combined} shows the convergence rates of Algorithm~\ref{alg_MO} (left) and Algorithm~\ref{alg_beamforming} (right) for different number of BS antennas, $M=\{4, 8, 12, 16\}$. For Algorithm~\ref{alg_MO}, it plots the gradient of the objective function, i.e., $\|{\rm{grad}}_{\tilde{\q{V}}_{l,r}} f(\tilde{\q{V}}_l)\|_2$, as a function of the number of iterations. It is observed that the gradient evolves across iterations for varying numbers of antennas. Initially, the gradient norm declines rapidly regardless of the number of BS antennas. Thus, Algorithm~\ref{alg_MO} quickly approaches the optimal regions with lower gradient norms. As the iterations progress, this reduction becomes more gradual, with frequent fluctuations. This indicates the algorithm's ability to adjust the step size and direction based on gradient guidance.

For Algorithm~\ref{alg_beamforming}, this figure illustrates the sensing sum rate as a function of the number of iterations. The algorithm is considered converged when the normalized objective function increases by less than \( \epsilon = \num{e-3} \). As shown in Fig.~\ref{fig_Convergence_combined}, the sum rate rises rapidly in the initial iterations before gradually saturating, demonstrating the algorithm’s fast convergence. Notably, it achieves convergence in fewer than five iterations, regardless of the number of BS antennas.

\subsection{Beampattern Gains}\label{sec_beamgain}
As an example, we consider four targets ($T=\num{4}$) with their directions from the  secondary BS to be  \qty{-40}{\degree}, \qty{-15}{\degree}, \qty{10}{\degree}, and \qty{35}{\degree}. The proposed system employs beamforming to enhance radar functionality by transmitting and receiving signals in specific directions, enabling precise target sensing through echo signal processing. Algorithm~\ref{alg_beamforming} facilitates beam formation and steering, improving signal quality, target detection, and interference mitigation \cite{Liu2018Radar, Zhenyao2023}.

The secondary BS transmitted signal, $\q{s}_l$, to illuminate targets, while the sensing combiners, i.e., $\q{u}_{l,t}$, are optimized for clear reception. The radar function is characterized by three key beampatterns: (i) $p_1(\theta) = | \q{a}_{l,t}^{\rm{H}} \q{s}_l^* |^2$ represents transmitted energy dispersion across angle $\theta$, (ii) $p_2(\theta) = | (\q{u}_{l,t}^*)^{\rm{H}} \q{b}_{l,t} |^2$ quantifies the system’s sensitivity to reflected energy across different angles, and (iii) $p_3(\theta) = | (\q{u}_{l,t}^*)^{\rm{H}} \q{b}_{l,t} \q{a}_{l,t}^{\rm{H}} \q{s}_l^* |^2$ integrates the effects of transmission and reflection for a complete representation \cite{Zargari2025}.

\begin{figure}[!t]\vspace{-2mm}
    \centering
    \includegraphics[width=0.45\textwidth]{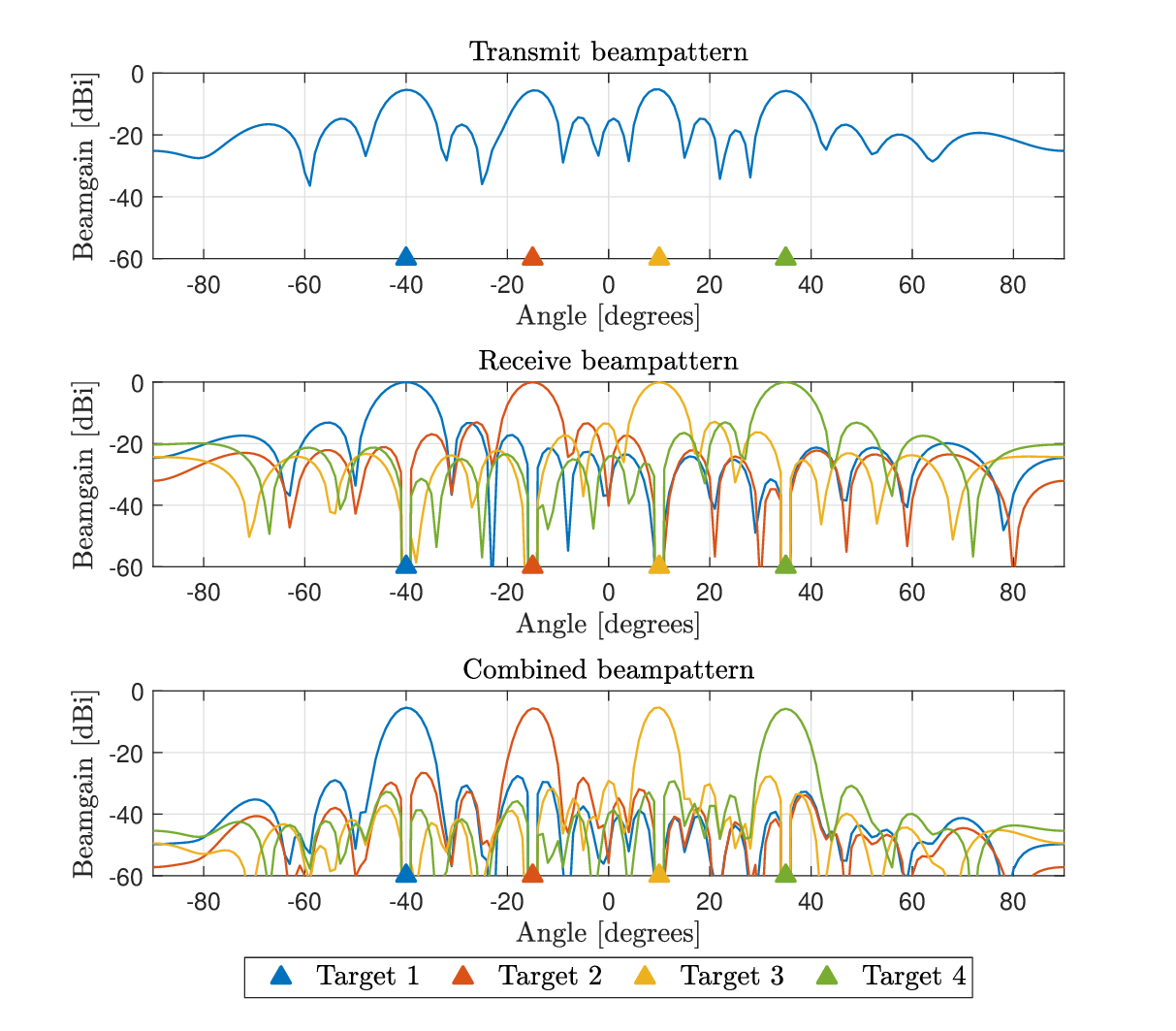}
    \vspace{-2mm}
    \caption{Beampatterns of the radar functionality of WB-cognitive scheme.}
    \label{fig_BeamGain_proposed} \vspace{-2mm}
\end{figure}

\begin{figure}[!t]\vspace{-2mm}
    \centering
    \includegraphics[width=0.45\textwidth]{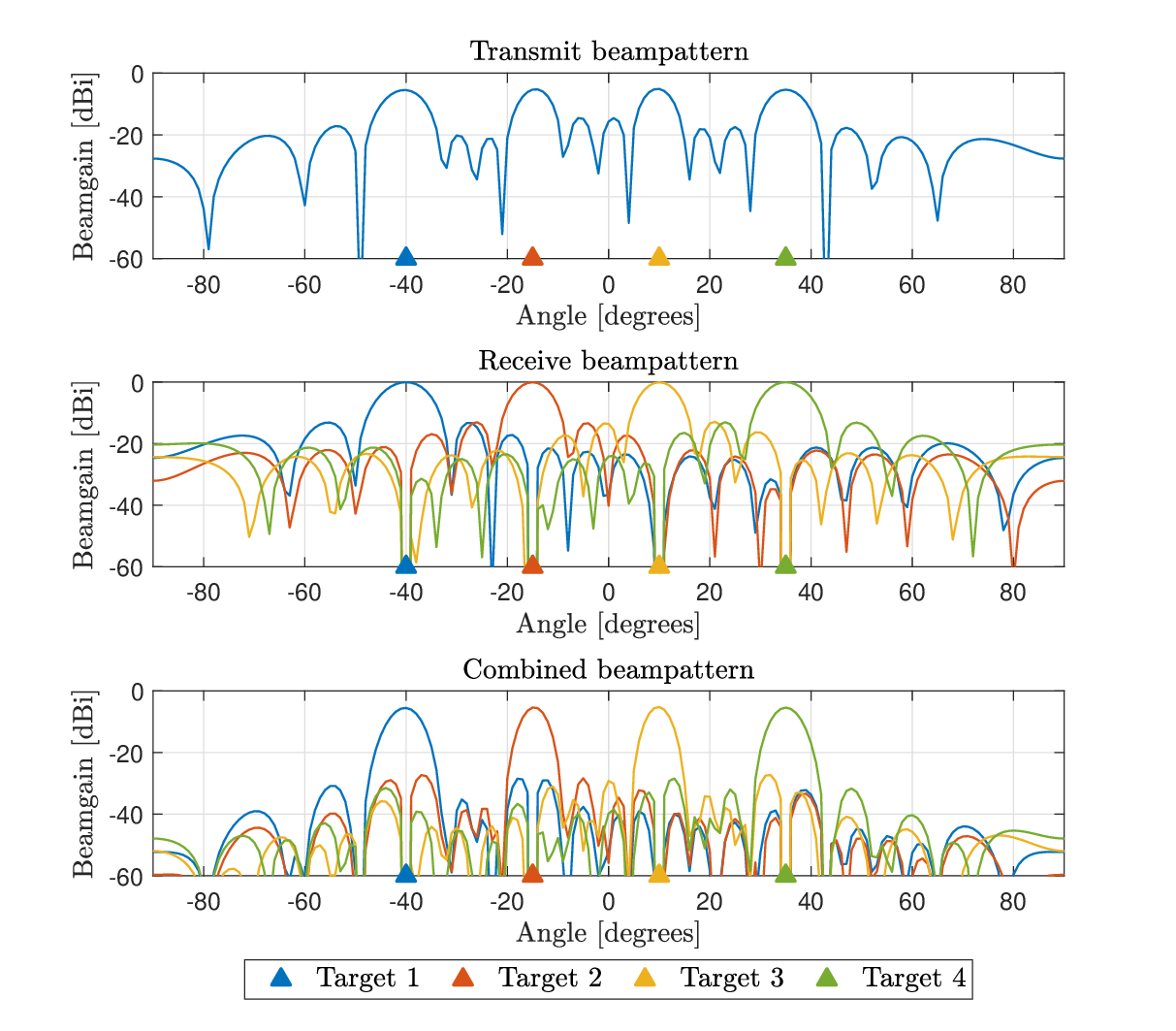}
    \vspace{-2mm}
    \caption{Beampatterns of the radar functionality of sensing-only scheme.}
    \label{fig_BeamGain_Sens} \vspace{-2mm}
\end{figure}

Fig.~\ref{fig_BeamGain_proposed} and Fig.~\ref{fig_BeamGain_Sens} plot $p_1(\theta), p_2(\theta),$ and $p_3(\theta) $ for the WB-cognitive scheme and sensing-only approach, respectively. Both schemes successfully identify target locations, as per the beampattern peaks.  However, the nulls and valleys of these patterns differ. The sensing-only scheme exhibits deeper nulls, resulting in a cleaner and more defined beampattern due to the absence of communication interference. In contrast, the WB-cognitive scheme has higher nulls. 

To further quantify this difference, Table~\ref{table_MSE_target} shows the MSE of target angle estimation. In addition to the WB-cognitive and sensing-only schemes, the non-cooperative scheme is also considered. The MSE is defined as ${\rm{MSE}} = \mathbb{E}\{(\theta - \hat{\theta})\}$, where $\hat{\theta}$ is the estimated value of the target direction $\theta$. The sensing-only approach achieves the lowest MSE, confirming its superior sensing accuracy in an interference-free environment. Due to the coexistence of primary communication signals, the WB-cognitive scheme has a slightly higher MSE than the sensing-only approach. Conversely, the non-cooperative scheme demonstrates the highest MSE among the three approaches. This is attributed to severe communication-sensing interference (non-cooperative operations), which disrupts both sensing and communication capabilities (Fig.~\ref{fig_CommRate_BSantennas} and Fig.~\ref{fig_SensRate_BSantennas}).

\begin{table}[!t]
\centering
\caption{MSE comparison of target angle estimation.}\vspace{-2mm}
\label{table_MSE_target}
\begin{tabular}{|l|lll|}
\hline
\multicolumn{1}{|c|}{\multirow{2}{*}{\textbf{Scheme}}} & \multicolumn{3}{c|}{\textbf{MSE}}                                                \\ \cline{2-4} 
\multicolumn{1}{|c|}{}                        & \multicolumn{1}{l|}{\textbf{Transmit}} & \multicolumn{1}{l|}{\textbf{Receive}} & \textbf{Combined} \\ \hline \hline
WB-cognitive                                  & \multicolumn{1}{l|}{\num{4.62e-5}}         & \multicolumn{1}{l|}{\num{3.70e-6}}       & \num{1.73e-5}         \\ \hline
Sensing-only                                & \multicolumn{1}{l|}{\num{3.61e-5}}         & \multicolumn{1}{l|}{\num{1.90e-6}}        &  \num{1.54e-5}         \\ \hline
Non-cooperative                                  & \multicolumn{1}{l|}{\num{1.74e-4}}         & \multicolumn{1}{l|}{\num{2.28e-5}}        &    \num{1.93e-5}      \\ \hline
\end{tabular}
\vspace{-2mm}
\end{table}

\subsection{Communication and Sensing Sum Rates}

\begin{figure}[!t]\vspace{-2mm}
    \centering
    \includegraphics[width=0.45\textwidth]{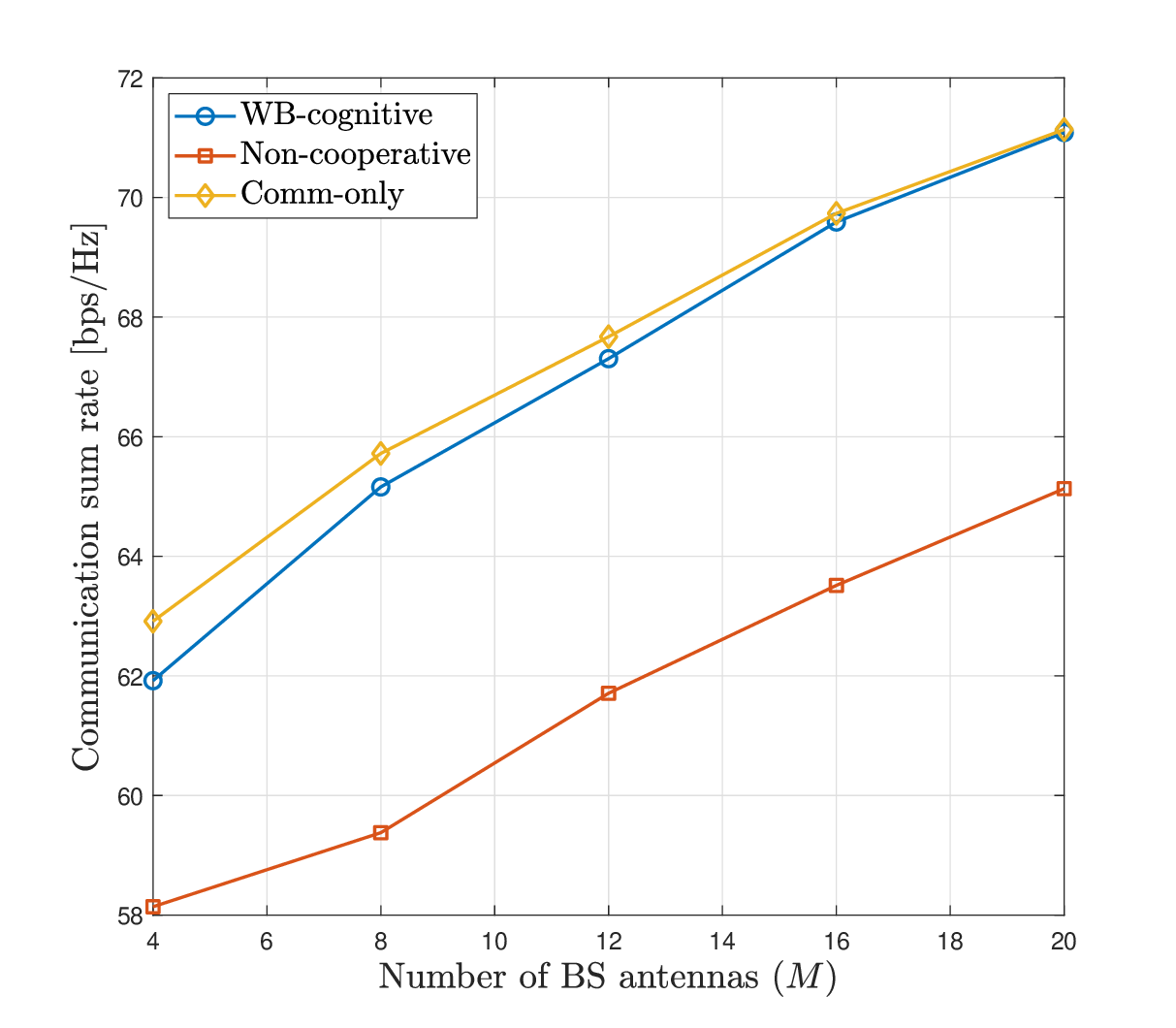}
    \vspace{-2mm}
    \caption{Communication sum rate as a function of the number of BS antennas.}
    \label{fig_CommRate_BSantennas} \vspace{-3mm}
\end{figure}

Fig.~\ref{fig_CommRate_BSantennas} compares the communication performance of the WB-cognitive, non-cooperative, and communication-only schemes as a function of the number of BS antennas, $M=N$. This figure shows that increasing the number of BS antennas improves the communication rate across all schemes. This is primarily due to a larger antenna array's enhanced spatial multiplexing capabilities.

\begin{figure}[!t]\vspace{-2mm}
    \centering
    \includegraphics[width=0.45\textwidth]{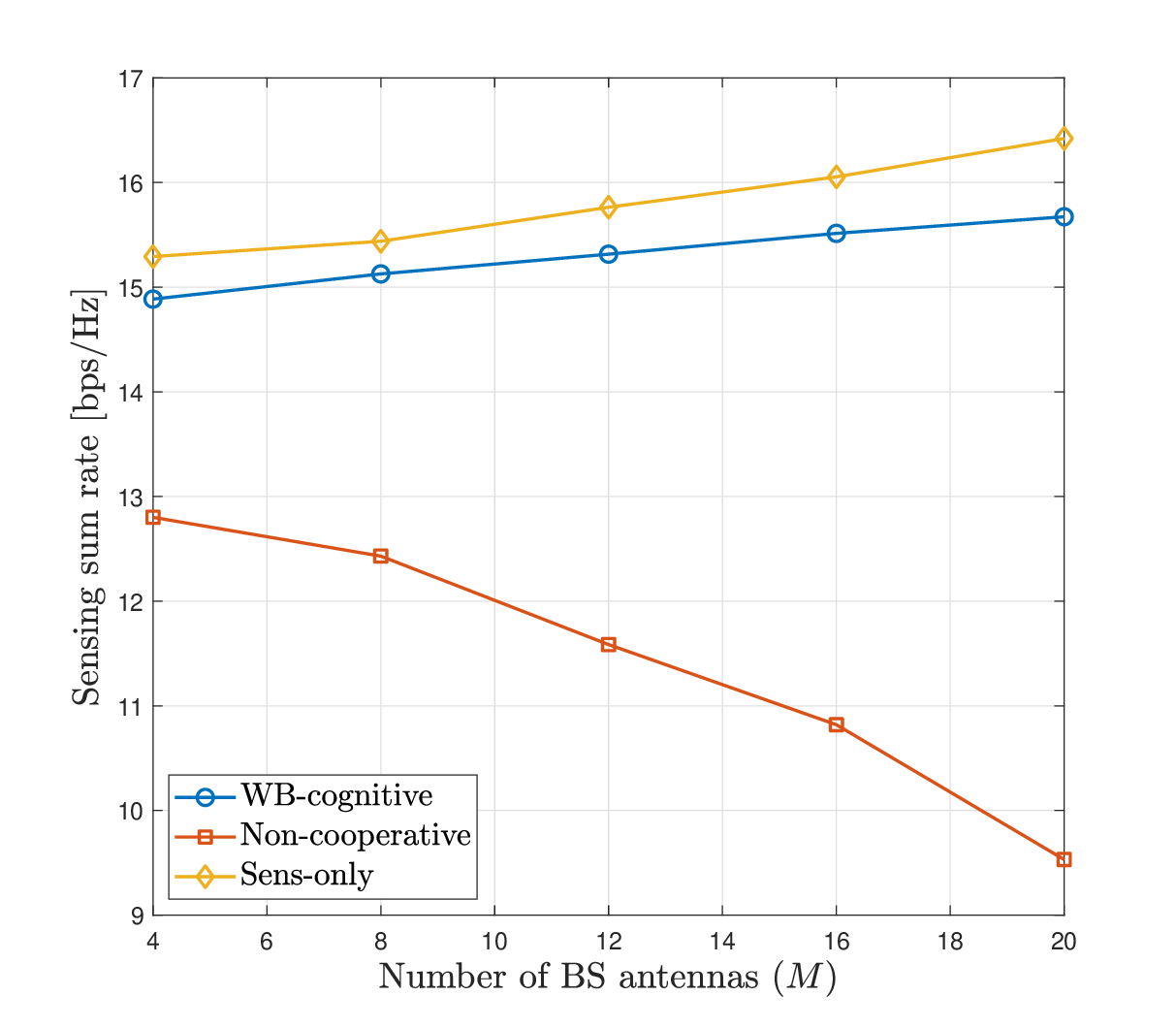}
    \vspace{-2mm}
    \caption{Sensing sum rate as a function of the number of BS antennas.}
    \label{fig_SensRate_BSantennas} \vspace{-2mm}
\end{figure}

Fig.~\ref{fig_CommRate_BSantennas} also illustrates the impact of secondary sensing interference on primary communication performance. The communication-only scheme achieves the highest sum rate as it operates without interference. The WB-cognitive beamforming design maintains a comparable communication sum rate while enabling sensing at the secondary BS (Fig.\ref{fig_SensRate_BSantennas}). In contrast, the non-cooperative design yields the lowest sum rate due to uncoordinated interference. For instance, with $M=\num{12}$, the WB-cognitive scheme achieves a \qty{10.0}{\percent} gain over the non-cooperative approach, highlighting the benefits of coordination.

Fig.~\ref{fig_SensRate_BSantennas} examines the sensing sum rate for the WB-cognitive, non-cooperative, and sensing-only schemes as a function of the number of BS antennas, $M=N$. This figure reflects the impact of inter-system interference on sensing performance, specifically the interference from primary communication on secondary sensing. Although a higher number of BS antennas enhances the sensing sum rate for the WB-cognitive and sensing-only schemes, it leads to a decline in the sensing sum rate for the non-cooperative scheme. This is because the WB-cognitive and sensing-only schemes benefit from the spatial multiplexing gains of a larger antenna array, whereas the non-cooperative scheme is hindered by increased communication interference at the secondary BS. Additionally, the WB-cognitive beamforming design effectively mitigates inter-system interference, leading to improved sensing (as well as communication (Fig.~\ref{fig_CommRate_BSantennas})) sum rate performance.  For example, with $M=\num{12}$, it achieves a \qty{32.2}{\percent} higher sensing sum rate compared to the non-cooperative scheme.

\subsection{Effects of Cognitive Operation}
\begin{figure}[!t]\vspace{-2mm}
    \centering
    \includegraphics[width=0.45\textwidth]{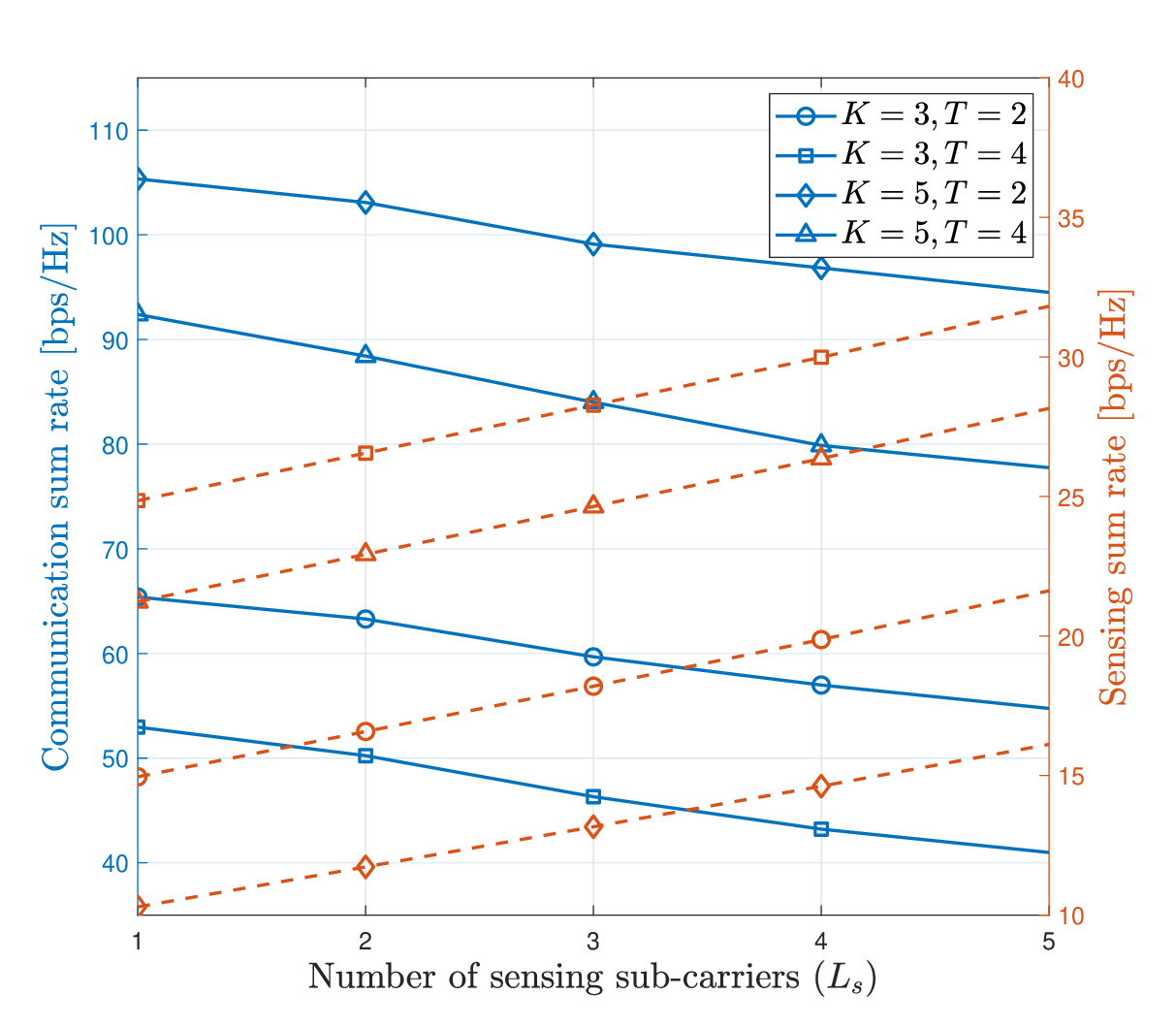}
    \vspace{-2mm}
    \caption{Communication and sensing sum rates as functions of the number of sensing sub-carriers ($L_s$). }
    \label{fig_SumRate_SubCarries} \vspace{-2mm}
\end{figure}

Fig.~\ref{fig_SumRate_SubCarries} examines the impact of the proposed WB-cognitive scheme on communication and sensing performance. It plots the communication sum rate (left \( y \)-axis) and sensing sum rate (right \( y \)-axis) as functions of the number of sub-carriers allocated for target sensing, \( L_s \).  

A key observation is that increasing \( L_s \) improves the sensing sum rate but reduces the communication sum rate. For instance, with \( K=3 \) and \( T=2 \), raising \( L_s \) from \num{1} to \num{4} enhances the sensing sum rate by \qty{32.88}{\percent} while decreasing the communication sum rate by \qty{12.86}{\percent}, primarily due to inter-system interference between communication and sensing.  

This highlights the trade-off in wideband radar-communication coexistence systems, where dynamic spectrum sharing is crucial. Unlike conventional systems, WB-cognitive approaches require strategic resource allocation to balance both functions. Prioritizing one inherently limits the other due to shared bandwidth and power constraints.  

Additionally, increasing the number of communication users \( K \) while keeping the number of sensing targets \( T \) constant improves the communication sum rate but degrades sensing performance, as additional users introduce more interference.

\subsection{CSI and SI Impairments}
\begin{figure}[!t]\vspace{-2mm}
    \centering
    \includegraphics[width=0.45\textwidth]{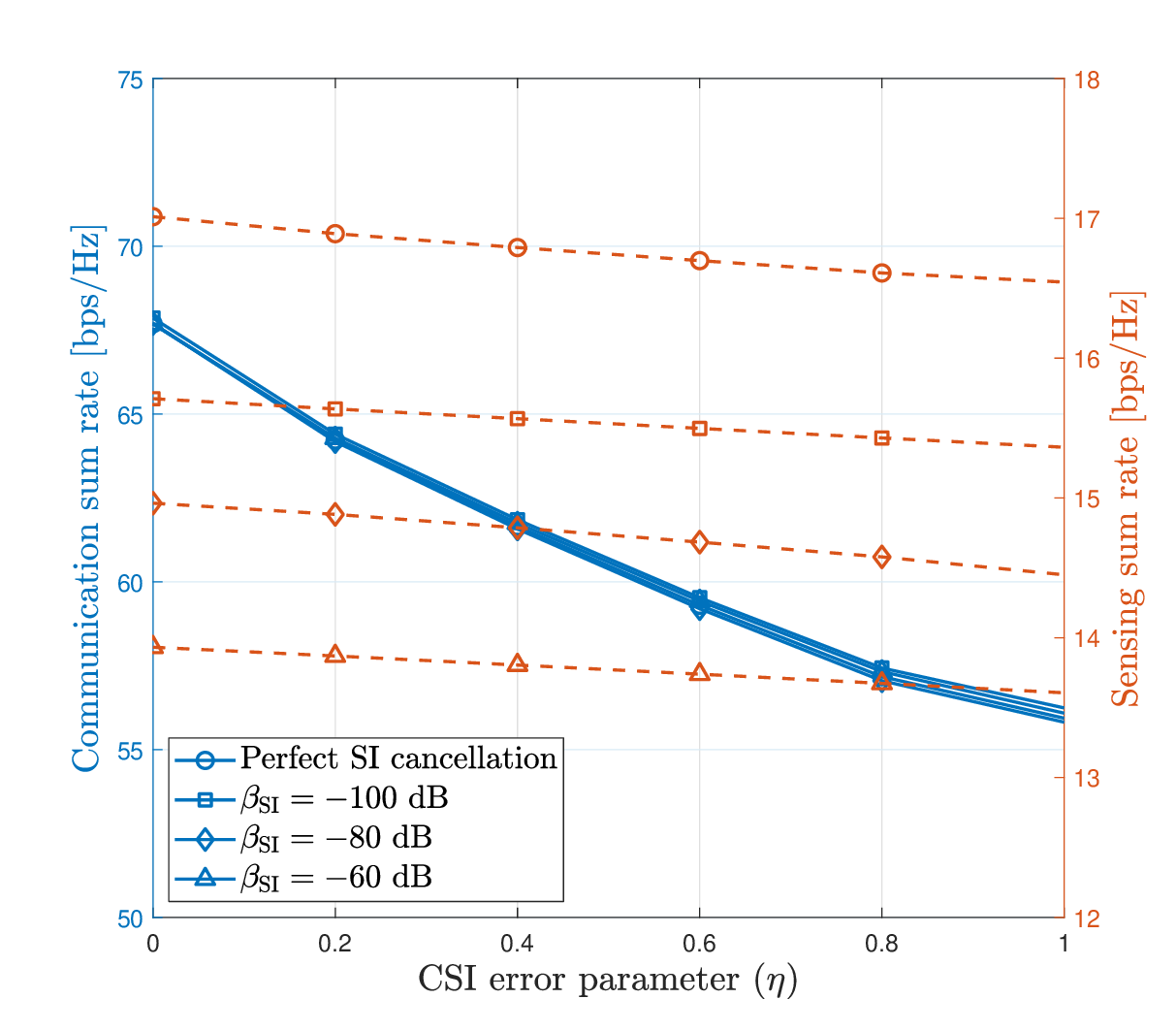}
    \vspace{-2mm}
    \caption{Communication and sensing sum rates as functions of CSI imperfection for various residual SI values. }
    \label{fig_ComAndSenRate_eta} \vspace{-2mm}
\end{figure}

The impacts of imperfect CSI and SI cancellation on communication and sensing sum rates are analyzed in Fig.~\ref{fig_ComAndSenRate_eta}. Specifically, in the WB-cognitive system, CSI errors compromise communication beamforming accuracy at the primary BS, meanwhile, SI cancellation errors, caused by incomplete suppression of SI, interfere with the reception of echo signals at the secondary BS \cite{Shengli2004, Diluka2023ImperfectCSI}. To model CSI errors, the true communication channel $x$ is represented as $\tilde{x} = x + e$, where $x \in \{[\q{h}_{l,k}]_m, [\q{g}_{l,k}]_n, [\q{F}_{l}]_{m,n}\}$ for $m \in \{1,\ldots, M\}$ and $n \in \{1,\ldots, N\}$. Here, $\tilde{x}$ is the estimated channel and $e$ is estimation noise distributed as $ e \sim \mathcal{N} (0, \sigma_e^2)$ \cite{KayEstimation1993, Shengli2004, Diluka2023ImperfectCSI}. The error variance, $\sigma_e^2$, is a key parameter that reflects the quality of channel estimation \cite{KayEstimation1993, Shengli2004, Diluka2023ImperfectCSI}. It can be modeled as $\sigma_e^2  = \eta |x|^2$, where $|x|$ is the magnitude of the true channel value and $0\leq \eta \leq 1$. Thus, $\eta$ measures the magnitude of CSI errors.

Fig.~\ref{fig_ComAndSenRate_eta} shows the relationships between the communication and sensing sum rates and the CSI error parameter, $\eta$. The communication sum rate (i.e., left $y$-axis) is highly sensitive to CSI errors. In particular, it improves with better CSI estimation, i.e., as $\eta \rightarrow 1$ (perfect CSI), the communication sum rate increases. Conversely, it declines as $\eta$ deviates from $1$ due to mismatches between the actual and estimated channels. Additionally, the communication sum rate is unaffected by SI cancellation capacity as SI occurs only at the secondary BS.

Fig.~\ref{fig_ComAndSenRate_eta} also illustrates the sensing sum rate (i.e., right $y$-axis) as a function of $\eta$. Clearly,  while the sensing sum rate is less affected by CSI errors, it is more susceptible to imperfect SI cancellation. In particular, lower residual SI leads to a higher sensing sum rate, as the reduced interference allows the secondary BS to process echo signals more effectively. This underscores the importance of robust SI cancellation techniques in enhancing sensing performance. 

\section{Conclusion}While communication and radar coexistence in higher frequency bands is possible via CR techniques, only single-user, single-target, or narrowband CR systems have been studied. Thus, this study develops a wideband CR communication and sensing system supporting multiple users and targets. Sub-carrier allocation for communication and Sub-carrier selection for sensing are developed. Based on that, transmit beamforming at the primary BS and sensing signal design with combiners at the secondary radar BS are optimized to maximize the communication sum rate while ensuring sensing requirements, minimizing interference, and adhering to BS power constraints. As this problem is non-convex, an MO algorithm is developed for communication-only sub-carriers, while an AO algorithm using the generalized Rayleigh quotient and SDR techniques optimizes joint communication-sensing sub-carriers. The proposed approach significantly outperforms non-cooperative systems, laying the foundation for future advancements in wideband radar-communication coexistence.

Moving forward, several key research directions remain open. One crucial aspect is the development of advanced channel estimation techniques to enhance sensing accuracy and communication reliability in dynamic environments. Additionally, more sophisticated sub-carrier allocation strategies, potentially employing machine learning techniques, could enable real-time adaptability to changing spectrum conditions. Investigating spectrum sensing performance, including metrics such as probability of detection and false alarm rate, will be essential to quantify the system’s robustness against interference and environmental variations. Moreover, scalability to large-scale networks with distributed nodes, energy-efficient optimization techniques, and robust coexistence strategies under dynamic interference conditions could be explored. 

\appendices

\section{Proof of the Convergence of Algorithm \ref{alg_MO}}\label{apdx_proof_Al_1}
The complete proof of the convergence to a global minimizer can be found in \cite[Proposition 1]{zargari2024riemannian}. It shows that the limit point of the sequence generated by MO satisfies the global optimality criteria for $(\mathcal{P}5)$. In the following, we briefly outline the proof.
\begin{itemize}
    \item \textit{Boundedness and non-increasing property}: The sequence $\{f(\tilde{\q{V}}_{l, r})\}$ is monotonically non-increasing and bounded below. Since $\epsilon_r \to 0$, we have $f(\tilde{\q{V}}_{l, r+1}) \leq f(\tilde{\q{V}}_{l, r})$, implying convergence of the sequence.
    \item \textit{Convergence to a limit point}:
    By the Bolzano-Weierstrass theorem, the sequence $\{\tilde{\q{V}}_r\}$, being bounded, has a convergent subsequence. Let $\tilde{\q{V}}^*$ be the limit of this subsequence, and $\tilde{\q{V}}^*$ satisfies the condition, $f(\tilde{\q{V}}_l^*) = \min_{\tilde{\q{V}}_l} ~f(\tilde{\q{V}}_{l, r})$.
    \item \textit{Global optimality}
    To prove global optimality, note that for any feasible$\tilde{\q{V}}$, $f(\tilde{\q{V}}_l^*) \leq f(\tilde{\q{V}}_l)$. Therefore, $\tilde{\q{V}}_l^*$ is a global minimizer.
\end{itemize}
This completes the proof.

\bibliographystyle{IEEEtran}
\bibliography{IEEEabrv,ref}

\begin{thebibliography}{10}
\providecommand{\url}[1]{#1}
\csname url@samestyle\endcsname
\providecommand{\newblock}{\relax}
\providecommand{\bibinfo}[2]{#2}
\providecommand{\BIBentrySTDinterwordspacing}{\spaceskip=0pt\relax}
\providecommand{\BIBentryALTinterwordstretchfactor}{4}
\providecommand{\BIBentryALTinterwordspacing}{\spaceskip=\fontdimen2\font plus
\BIBentryALTinterwordstretchfactor\fontdimen3\font minus
  \fontdimen4\font\relax}
\providecommand{\BIBforeignlanguage}[2]{{%
\expandafter\ifx\csname l@#1\endcsname\relax
\typeout{** WARNING: IEEEtran.bst: No hyphenation pattern has been}%
\typeout{** loaded for the language `#1'. Using the pattern for}%
\typeout{** the default language instead.}%
\else
\language=\csname l@#1\endcsname
\fi
#2}}
\providecommand{\BIBdecl}{\relax}
\BIBdecl

\bibitem{Azar2024}
A.~Hakimi, D.~Galappaththige, and C.~Tellambura, ``A roadmap for {NF-ISAC} in
  {6G}: A comprehensive overview and tutorial,'' \emph{Entropy}, vol.~26,
  no.~9, Sept. 2024.

\bibitem{Khawar2014}
A.~Khawar, A.~Abdel-Hadi, and T.~C. Clancy, ``Spectrum sharing between {S}-band
  radar and {LTE} cellular system: A spatial approach,'' in \emph{Proc. IEEE
  Int. Symp. Dynamic Spectr. Access Netw.}, Apr. 2014, pp. 7--14.

\bibitem{Wang2019}
T.~Wang, J.~Huang, X.~Wang, S.~Wang, and Z.~Fei, ``Sum-rate optimization for
  radar and communication coexistence,'' in \emph{Proc. 11th Int. Conf.
  Wireless Commun. Signal Process.}, Oct. 2019, pp. 1--5.

\bibitem{Qian2021}
J.~Qian, A.~Zhang, L.~Zhao, and F.~Jia, ``An iterative algorithm design for
  radar-communication coexistence,'' in \emph{Proc. Int. Conf. Control, Autom
  Inf. Sciences}, Oct. 2021, pp. 336--340.

\bibitem{Mao2024}
H.~Mao, H.~Zhu, Y.~He, G.~Yu, and R.~Yin, ``A framework for co-existence of
  radar and communication with joint design,'' \emph{EURASIP J. Wirel. Commun.
  Netw.}, vol. 2024, no.~1, Jul 2024.

\bibitem{xu2024}
Y.~Xu, Y.~Li, and T.~Q.~S. Quek, ``{RIS}-enhanced cognitive integrated sensing
  and communication: Joint beamforming and spectrum sensing,'' \emph{arXiv},
  2024.

\bibitem{Liu2024}
M.~Liu \emph{et~al.}, ``Joint beamforming design for double active
  {RIS}-assisted radar-communication coexistence systems,'' \emph{{IEEE} Trans.
  on Cogn. Commun. Netw.}, vol.~10, no.~5, pp. 1704--1717, Oct. 2024.

\bibitem{Elfiatoure2024}
M.~Elfiatoure, M.~Mohammadi, H.~Q. Ngo, P.~J. Smith, and M.~Matthaiou,
  ``Protecting massive {MIMO}-radar coexistence: Precoding design and power
  control,'' \emph{{IEEE} Open J. Commun. Soc.}, vol.~5, pp. 276--293, Jan.
  2024.

\bibitem{Zheng2019}
L.~Zheng, M.~Lops, Y.~C. Eldar, and X.~Wang, ``Radar and communication
  coexistence: An overview: A review of recent methods,'' \emph{{IEEE} Signal
  Process. Mag.}, vol.~36, no.~5, pp. 85--99, Sept. 2019.

\bibitem{Hilal2023}
W.~Hilal, S.~A. Gadsden, and J.~Yawney, ``Cognitive dynamic systems: A review
  of theory, applications, and recent advances,'' \emph{Proc. {IEEE}}, vol.
  111, no.~6, pp. 575--622, Jun. 2023.

\bibitem{Rappaport2015}
T.~S. Rappaport, G.~R. MacCartney, M.~K. Samimi, and S.~Sun, ``Wideband
  millimeter-wave propagation measurements and channel models for future
  wireless communication system design,'' \emph{{IEEE} Trans. Commun.},
  vol.~63, no.~9, pp. 3029--3056, Sept. 2015.

\bibitem{Zhenyao2023}
Z.~He \emph{et~al.}, ``Full-duplex communication for {ISAC}: Joint beamforming
  and power optimization,'' \emph{{IEEE} J. Sel. Areas Commun.}, vol.~41,
  no.~9, pp. 2920--2936, Sept. 2023.

\bibitem{Uwaechia2020}
A.~N. Uwaechia and N.~M. Mahyuddin, ``A comprehensive survey on millimeter wave
  communications for fifth-generation wireless networks: Feasibility and
  challenges,'' \emph{{IEEE} Access}, vol.~8, pp. 62\,367--62\,414, Apr. 2020.

\bibitem{Mohammadi2023}
M.~Mohammadi, Z.~Mobini, D.~Galappaththige, and C.~Tellambura, ``A
  comprehensive survey on full-duplex communication: Current solutions, future
  trends, and open issues,'' \emph{{IEEE} Commun. Surveys Tuts.}, vol.~25,
  no.~4, pp. 2190--2244, 4th Quart. 2023.

\bibitem{Diluka2024CFFD}
D.~Galappaththige, M.~Mohammadi, H.~Q. Ngo, M.~Matthaiou, and C.~Tellambura,
  ``Cell-free full-duplex communication -- {An} overview,'' \emph{arXiv}, 2024.

\bibitem{tse_viswanath_2005}
D.~Tse and P.~Viswanath, \emph{Fundamentals of Wireless Communication}.\hskip
  1em plus 0.5em minus 0.4em\relax Cambridge University Press, 2005.

\bibitem{Marzettabook2016}
T.~L. Marzetta, E.~G. Larsson, H.~Yang, and H.~Q. Ngo, \emph{Fundamentals of
  Massive {MIMO}}.\hskip 1em plus 0.5em minus 0.4em\relax Cambridge Univ.
  Press, 2016.

\bibitem{Nayebi2018}
E.~Nayebi and B.~D. Rao, ``Semi-blind channel estimation for multiuser massive
  {MIMO} systems,'' \emph{{IEEE} Trans. Signal Process.}, vol.~66, no.~2, pp.
  540--553, Jan. 2018.

\bibitem{Tsinos2021Joint}
C.~G. Tsinos, A.~Arora, S.~Chatzinotas, and B.~Ottersten, ``Joint transmit
  waveform and receive filter design for dual-function radar-communication
  systems,'' \emph{{IEEE} J. Sel. Topics Signal Process.}, vol.~15, no.~6, pp.
  1378--1392, Nov. 2021.

\bibitem{Wu2018}
L.~Wu, P.~Babu, and D.~P. Palomar, ``Transmit waveform/receive filter design
  for {MIMO} radar with multiple waveform constraints,'' \emph{{IEEE} Trans.
  Signal Process.}, vol.~66, no.~6, pp. 1526--1540, Mar. 2018.

\bibitem{positioningLTE}
``Positioning techniques for mobile devices in {LTE},'' Jul. 2015. Available
  Online:
  \url{https://www.hsc.com/resources/blog/positioning-techniques-for-mobile-devices-in-lte/}.

\bibitem{Liu2010}
T.~Liu, C.~Yang, and L.-L. Yang, ``A low-complexity subcarrier-power allocation
  scheme for frequency-division multiple-access systems,'' \emph{{IEEE} Trans.
  Wireless Commun.}, vol.~9, no.~5, pp. 1564--1570, May 2010.

\bibitem{Gouissem2016}
A.~Gouissem, R.~Hamila, N.~Al-Dhahir, and S.~Foufou, ``Sparsity-aware
  narrowband interference mitigation and subcarriers selection in {OFDM}-based
  cognitive radio networks,'' in \emph{Proc. IEEE 84th Veh. Technol. Conf.},
  Sept. 2016, pp. 1--7.

\bibitem{Liu2022}
F.~Liu, Y.-F. Liu, A.~Li, C.~Masouros, and Y.~C. Eldar, ``Cram\'{e}r-{Rao}
  bound optimization for joint radar-communication beamforming,'' \emph{{IEEE}
  Trans. Signal Process.}, vol.~70, pp. 240--253, Jan. 2022.

\bibitem{James2010RadarBook}
M.~A. Richards, J.~A. Scheer, and W.~A. Holm, Eds., \emph{Principles of Modern
  Radar: Basic principles}, ser. Radar, Sonar and Navigation.\hskip 1em plus
  0.5em minus 0.4em\relax Institution of Engineering and Technology, 2010.

\bibitem{Stoica2007}
P.~Stoica, J.~Li, and Y.~Xie, ``On probing signal design for {MIMO} radar,''
  \emph{{IEEE} Trans. Signal Process.}, vol.~55, no.~8, pp. 4151--4161, Jul.
  2007.

\bibitem{Cui2014}
G.~Cui, H.~Li, and M.~Rangaswamy, ``{MIMO} radar waveform design with constant
  modulus and similarity constraints,'' \emph{{IEEE} Trans. Signal Process.},
  vol.~62, no.~2, pp. 343--353, Jan. 2014.

\bibitem{Tang2019}
B.~Tang and J.~Li, ``Spectrally constrained {MIMO} radar waveform design based
  on mutual information,'' \emph{{IEEE} Trans. Signal Process.}, vol.~67,
  no.~3, pp. 821--834, Feb. 2019.

\bibitem{zargari2024riemannian}
S.~Zargari, D.~Galappaththige, C.~Tellambura, and H.~Vincent~Poor, ``A
  {Riemannian} manifold approach to constrained resource allocation in
  {ISAC},'' \emph{{IEEE} Trans. Commun.}, pp. 1--1, 2024.

\bibitem{zargari2024CFISAC}
S.~Zargari, D.~Galappaththige, C.~Tellambura, and G.~Y. Li, ``Downlink
  beamforming for cell-free {ISAC}: A fast complex oblique manifold approach,''
  \emph{{IEEE} Trans. Wireless Commun.}, 2025.

\bibitem{Shen2018FPpart2}
K.~Shen and W.~Yu, ``Fractional programming for communication systems—part
  {II}: Uplink scheduling via matching,'' \emph{{IEEE} Trans. Signal Process.},
  vol.~66, no.~10, pp. 2631--2644, May 2018.

\bibitem{Shen2018}
------, ``Fractional programming for communication systems—part {I}: Power
  control and beamforming,'' \emph{{IEEE} Trans. Signal Process.}, vol.~66,
  no.~10, pp. 2616--2630, May 2018.

\bibitem{boyd2004convex}
S.~Boyd and L.~Vandenberghe, \emph{Convex {O}ptimization}.\hskip 1em plus 0.5em
  minus 0.4em\relax Cambridge, U.K.: Cambridge Univ. Press, 2004.

\bibitem{liu2020simple}
C.~Liu and N.~Boumal, ``Simple algorithms for optimization on {Riemannian}
  manifolds with constraints,'' \emph{Appl. Math. Optim.}, vol.~82, pp.
  949--981, Mar. 2020.

\bibitem{Shewchuk1994}
\BIBentryALTinterwordspacing
J.~R. Shewchuk, ``An introduction to the conjugate gradient method without the
  agonizing pain,'' USA, Tech. Rep., 1994. [Online]. Available:
  \url{http://www.cs.cmu.edu/~quake-papers/painless-conjugate-gradient.pdf}
\BIBentrySTDinterwordspacing

\bibitem{bezdek2003convergence}
J.~C. Bezdek and R.~J. Hathaway, ``Convergence of alternating optimization,''
  \emph{Neural, Parallel \& Scientific Computations}, vol.~11, no.~4, pp.
  351--368, Dec. 2003.

\bibitem{Stanczak2008book}
S.~Stanczak, \emph{\BIBforeignlanguage{eng}{Fundamentals of Resource Allocation
  in Wireless Networks Theory and Algorithms}}, 2nd~ed.\hskip 1em plus 0.5em
  minus 0.4em\relax Berlin, Heidelberg: Springer Berlin Heidelberg, 2008.

\bibitem{Luo2010}
Z.-q. Luo, W.-k. Ma, A.~M.-c. So, Y.~Ye, and S.~Zhang, ``Semidefinite
  relaxation of quadratic optimization problems,'' \emph{{IEEE} Signal Process.
  Mag.}, vol.~27, no.~3, pp. 20--34, May 2010.

\bibitem{polik2010interior}
I.~P’olik and T.~Terlaky, \emph{Interior Point Methods for Nonlinear
  Optimization}.\hskip 1em plus 0.5em minus 0.4em\relax Berlin, Germany; New
  York, NY, USA: Springer, 2010.

\bibitem{so2007approximating}
A.~M.~C. So, J.~Zhang, and Y.~Ye, ``On approximating complex quadratic
  optimization problems via semidefinite programming relaxations,'' \emph{Math.
  Program.}, vol. 110, no.~1, pp. 93--110, Jun. 2007.

\bibitem{3GPP2024}
\BIBentryALTinterwordspacing
``{3GPP TR} 38.901, {Study} on channel model for frequencies from 0.5 to 100
  {GHz}, {V}.18.0.0 {R}el. 18,'' Apr. 2024. [Online]. Available:
  \url{https://portal.3gpp.org/desktopmodules/Specifications/SpecificationDetails.aspx?specificationId=3173}
\BIBentrySTDinterwordspacing

\bibitem{Liu2018Radar}
F.~Liu, C.~Masouros, A.~Li, H.~Sun, and L.~Hanzo, ``{MU-MIMO} communications
  with {MIMO} radar: From co-existence to joint transmission,'' \emph{{IEEE}
  Trans. Wireless Commun.}, vol.~17, no.~4, pp. 2755--2770, Apr. 2018.

\bibitem{Zargari2025}
S.~Zargari, D.~Galappaththige, and C.~Tellambura, ``Transmit power-efficient
  beamforming design for integrated sensing and backscatter communication,''
  \emph{{IEEE} Open J. Commun. Soc.}, vol.~6, pp. 775--792, Jan. 2025.

\bibitem{Shengli2004}
S.~Zhou and G.~Giannakis, ``How accurate channel prediction needs to be for
  transmit-beamforming with adaptive modulation over {Rayleigh} {MIMO}
  channels?'' \emph{{IEEE} Trans. Wireless Commun.}, vol.~3, no.~4, pp.
  1285--1294, Jul. 2004.

\bibitem{Diluka2023ImperfectCSI}
D.~Galappaththige, F.~Rezaei, C.~Tellambura, and S.~Herath, ``Beamforming
  designs for enabling symbiotic backcom multiple access under imperfect
  {CSI},'' \emph{{IEEE} Access}, vol.~11, pp. 89\,986--90\,005, Aug. 2023.

\bibitem{KayEstimation1993}
S.~M. Kay, \emph{Fundamentals of Statistical Signal Processing: Estimation
  Theory}.\hskip 1em plus 0.5em minus 0.4em\relax USA: Prentice-Hall, Inc.,
  1993.

\end{thebibliography}

\end{document}